\documentclass[a4paper]{article} %11pt,a4paper,letterpaper,twocolumn 
\usepackage{amsmath, amssymb, amsthm}
\usepackage[mathscr]{euscript}     %usage: \mathscr{ABC}
\usepackage{dsfont} % usage: \mathds{ABC} for nicer blackboard font
\usepackage[usenames,dvipsnames]{color}
\usepackage{graphicx} % graphics
\usepackage{underscore} % need it for a doi which includes an underscore "_"
\usepackage[shortcuts]{extdash} % for hyphenation of words after dashes 
\usepackage{authblk} % author / affil commae.g., \appendixpage
\usepackage{appendix} % appendix commands, e.g., \appendixpage
\usepackage{microtype} % improves line spacing with pdflatex
\usepackage{enumitem} % more itemize commands, in particular for spacing.
\setenumerate[1]{label=\textbf{\arabic*.}, ref=\arabic*}
\setenumerate[2]{label=\theenumi.\arabic*., ref=\theenumi.\arabic*}
\setenumerate[3]{label=\theenumi.\theenumii.\arabic*., ref=\theenumi.\theenumii.\arabic*}
\usepackage[margin=10pt,font=small,labelfont=bf,labelsep=endash,hypcap=true]{caption}
\usepackage[margin=10pt,font=small,labelfont=bf,labelformat=parens,labelsep=space,hypcap=false]{subcaption}
\usepackage{tikz} % inline graphics.
\usetikzlibrary{fit} % \usetikzlibrary{arrows,fit,positioning,shapes}
\usepackage{bbm}
\usepackage[linktoc=all]{hyperref} 
\usepackage[open,numbered]{bookmark} % additional pdfbookmark
\usepackage{breakurl} % needed only for latex, not pdflatex. Doesn't
\usepackage{algorithm}
\floatname{algorithm}{Protocol}
\usepackage{complexity}
\def\EQ#1{\begin{eqnarray}#1\end{eqnarray}}
\makeatletter
\newcommand{\vast}{\bBigg@{3}}
\newcommand{\Vast}{\bBigg@{6.5}}
\makeatother
\newcommand{\dm}[1]{\ket{#1}\bra{#1}}
\newcommand{\djj}{d\kern-0.4em\char"16\kern-0.1em}

\newcommand{\todo}[1]{{\color{red} #1}}

\newcommand{\ctrlZ}{\text{ctrl-}Z}
\newcommand{\aR}{\mathscr{R}}
\newcommand{\aS}{\mathscr{S}}

\usepackage{geometry}
\def\ith#1{#1^{th}}
\newtheorem{prop}{Proposition}\def\PRO{\begin{prop}}\def\ORP{\end{prop}}
\newtheorem{coro}{Corollary}\def\COR{\begin{coro}}\def\ROC{\end{coro}}
\newtheorem{theo}{Theorem}\def\TH{\begin{theo}}\def\HT{\end{theo}}
\def\TH{\begin{theo}}\def\HT{\end{theo}}
\newtheorem{defi}[prop]{Definition}\def\DE{\begin{defi}}\def\ED{\end{defi}}
\newtheorem{lemme}[prop]{Lemma}\def\LE{\begin{lemme}}\def\EL{\end{lemme}}
\def\eps{\epsilon}
\def\vec#1{{\bf #1}}
\def\tr{\text{Tr}}
\def\ora#1{\mathbf{#1}}
\def\bb#1{ \boxed{#1} }
\def\Id{\mathbbmss{1}}
\def\lo#1{\mathcal{L}(\mathcal{H}_{#1})}
\newcommand{\df}{\mathrel{\mathop:}=}
\input{Qcircuit.tex}

\title{Blind quantum computing with two almost identical states}

\author[1]{Vedran~Dunjko\footnote{\href{mailto:vedran.dunjko@uibk.ac.at}{vedran.dunjko@uibk.ac.at}}}
\author[2,3]{Elham Kashefi}

\affil[1]{Institute for Theoretical
    Physics, University of Innsbruck, Technikerstra{\ss}e 21, A-6020
    Innsbruck, Austria. }
\affil[2]{School of Informatics, University of Edinburgh,
10 Crichton Street, Edinburgh EH8 9AB, United Kingdom}
  \affil[3]{CNRS LTCI, Departement Informatique et Reseaux,
Telecom ParisTech, Paris CEDEX 13, France.}
\date{\today}

\begin{document}
%\date{\today}
%\title{Blind quantum computing with two almost identical states}
%\date{\today}
%\author{Vedran Dunjko}
%\email{vedran.dunjko@uibk.ac.at}
%\address{Institut f\"{u}r Theoretische Physik, Universit{\"{a}}t Innsbruck, Technikerstra{\ss}e 25, A-6020 Innsbruck, Austria}
%\author{Elham Kashefi}
%\address{
%School of Informatics, University of Edinburgh,
%10 Crichton Street, Edinburgh EH8 9AB, United Kingdom}
%\address{CNRS LTCI, Departement Informatique et Reseaux,
%Telecom ParisTech, Paris CEDEX 13, France}
\maketitle

\begin{abstract}
The question of whether a fully classical client can delegate a quantum computation to an untrusted quantum server while fully maintaining privacy (blindness) is one of the big open questions in quantum cryptography.
Both yes and no answers have important practical and theoretical consequences, and the question seems genuinely hard.
The state-of-the-art approaches to securely delegating quantum computation, without exception, rely on granting the client modest quantum powers, or on additional, non-communicating, quantum servers. In this work, we consider the single server setting, and push the boundaries of the minimal devices of the client, which still allow for blind quantum computation.
Our approach is based on the observation that, in many blind quantum computing protocols, the ``quantum" part of the protocol, from the clients perspective, boils down to the establishing classical-quantum correlations (independent from the computation) between the client and the server, following which the steering of the computation itself requires only classical communication. 

Here, we abstract this initial preparation phase, specifically for the Universal Blind Quantum Computation protocol of Broadbent, Fitzsimons and Kashefi. We identify sufficient criteria on the powers of the client, which still allow for secure blind quantum computation. We work in a universally composable framework, and provide a series of protocols, where each step reduces the number of differing states the client needs to be able to prepare.
As the limit of such reductions, we show that the capacity to prepare just two pure states, which have an arbitrarily high overlap (thus are arbitrarily close to identical), suffices for efficient and secure blind quantum computation.
\end{abstract}

%\tableofcontents
\section{Introduction}
An important question in modern (quantum) cryptography asks whether a fully classical Alice can securely delegate a quantum computation to an untrusted server Bob. The main two flavours of security one is interested in are \emph{blindness} -- meaning Alice's computation remains private and hidden from Bob -- and \emph{verifiability}  --meaning Alice has a mechanism which ensures the declared output of the computation is indeed correct.
An answer to this question would have obvious practical consequences, but also a profound impact on computational complexity theory and cryptography.
To this day, no such protocol with a fully classical Alice, which allows delegating all quantum computations while maintaining either verifiability or blindness, has been found. Nonetheless, protocols exist in which one or both properties can be achieved, if Alice is allowed modest quantum powers, or by utilizing multiple quantum servers (which cannot communicate).
Over the course of recent years, the envelope of such minimal additional requirements has been pushed, both theoretically \cite{Chi05,AS06,BFK09,ABE10,FK12, DKL12,Mor12,RUV13,MPDF13,Mor14,GMMR13,DFPR14,MDK15,MF12,BGS13,FBSYLPJR14,PDF15}, but also experimentally \cite{BKBFZW12,BFKW13}. The typical improvements of interest include minimizing the size of the overall quantum and/or classical communication, and simplifying Alice's required quantum devices, or the guarantees placed on them (in e.g. device-independent constructions).

In this work, we focus on the blindness aspects of delegating quantum computation. While the problem of classical verifiable delegated computation has well-known deep ties with complexity theory, it is perhaps less widely known 
that the same holds for the big question of a fully classical-Alice blind delegated quantum computing protocol as well.

 In particular, both existance, and non-existance of perfectly blind classical-client delegated computing protocols has highly non-trivial consequences in complexity theory, which may elucidate why finding such protocols (or proving they cannot exist) should be a hard task. These results, sketched out in this work, motivate us to approach the big question indirectly, and consider the problem of minimizing the quantum powers of Alice, while maintaining blindness in delegated quantum computation. 

In our approach, we focus on the family of protocols originating from the Universal Blind Quantum Computation (UBQC) protocol of Broadbent, Fitzsimons and Kashefi \cite{BFK09}.
Common to all these protocols is that the client first establishes a particular type of classical-quantum (CQ) correlated state, either by utilizing its own powers, or through the help of a second server, entangled, but not communicating with the first \cite{BFK09}. Following this, the entire computation proceeds through just classical communication. 

Here, we abstract this initial distribution phase as a fully defined functionality (in the sense used in composable cryptography), in which case the analysis of minimal powers needed for the rest of the protocol, reduces to the the minimal requirements for the implementation of this simpler functionality.
In the process, we identify the genuinely quantum properties of the UBQC protocol central to its security.

While our approach cannot yield a secure fully classical protocol, as such schemes which require quantum preparation, intuitively, must depend on some inherently quantum property, it does allow us to push the limits further.
In particular, in the UBQC protocol, Alice's device, roughly speaking, is capable of producing 8 distinct qubit states.
As a first step, we show that a weaker device, where Bob has more options to cheat, still suffices for perfect composable blindness.

Next, we give a protocol which utilizes an analogous device, capable of producing only 4 states, and can be securely used as a sub-routine, substituting the 8 state preparation device, while maintaining perfect security. 
This already implies that Alice capable of producing just 4 BB84-type states used in quantum key distribution~\cite{BB84} suffices. We note that 4 state protocols, which achieve differing flavours of security, have already been reported \cite{GMMR13, FBSYLPJR14}. 
Our approach, however, allows us to push the envelope further, and we give a protocol in which a device capable of preparing just two states (unitarily equivalent to $\ket{0}$ and $\ket{+} = 1/\sqrt{2}(\ket{0} + \ket{1})$ which can, again, securely be used as a substitute for the 4 state device. At this point, perfect security is no longer maintainable, but we achieve the next best thing - exponential security.
As the last step, we show how this generalizes, and we show that for any fidelity $F,$ no matter how large, there exist two states $\ket{\psi}, \ket{\phi},$ such that $F \leq  | \bra{\psi} \phi \rangle|^2,$ which can be used instead of the two states $\ket{0}, \ket{+},$ while maintaining exponential security.
In other words, we achieve blind quantum computation, where Alice's device can produce two pure states, which can be arbitrarily close to identical. Note that in the limit of fully identical states, the overall protocol would be fully classical.

\subsection{Paper overview}

For didactic purposes, we will start off by considering only the correlations which are established between Alice and Bob during a run of a protocol. In cryptographic language, the security definitions which arise from such analyses are ``local'' security definitions, and are not embedded in a wider framework of security which places guarantees on how the protocol performs when combined with other protocols. However, this will already suffice for us to place necessary criteria on what the powers of Alice must be, in this preparation phase.  
Later, we transform all the local results into a composable security framework, namely Abstract Cryptography \cite{MR11,Mau12, DFPR14,PMMRT15}, where all sufficiency statements are proven.

In Section \ref{prelim} we cover the preliminaries for this work: basic notation, definitions pertaining to delegated quantum computing protocols and a succinct description of the UBQC protocol, and we describe the basic ideas behind composable security, tailored for our setting.
Following this, in Section \ref{connections}, we sketch out some results regarding the implications the (non-)existance of a blind quantum computing would have in complexity theory.
 Next, in Section \ref{correlations} we study the necessary and sufficient correlations which must be established in the preparation phase of UBQC, for the protocol to be blind.
 In Section \ref{functionalities} we abstract the notion of preparation devices, and treat this preparation process from the perspective of composable security. We provide a couple of functionalities which are more general than assumed in UBQC (they allow for a broader variety of malevolent behaviour for a dishonest Bob), and prove they still suffice for a composably blind UBQC protocol. As an illustration, we show that the two-server protocol of \cite{BFK09} is perfectly composably secure.
 In the final Section \ref{minimal} we use the established formalism to prove that four, and then two states preparation (that is, a preparation in which Alice can generate only two differing states), still suffice for secure blind quantum computation. We conclude with a discussion section.

\section{Preliminaries}
\label{prelim}
\subsection{Basic notation}

Throughout the paper we will use $X/Z$ to denote the Pauli X/ Pauli Z gates, $H$ denotes the Hadamard gate,
$S$ denotes the phase gate such that $SS = Z$. The identity we denote with $\mathbbmss{1}.$
In general, with $\ket{\pm_\theta}$ we denote the state $\dfrac{1}{\sqrt{2}}( \ket{0} \pm e^{i\theta}\ket{1}),$ where $\theta$ is an angle.
With $\ctrlZ$ we dentoe the two-qubit controlled Z rotation: $\ctrlZ= \dm{0}\otimes \mathbbmss{1} + \dm{1}\otimes Z$.

The notation $A \approx_{\epsilon} B$, if $A$ and $B$ are states, indicates  $A$ and $B$ are $\epsilon$-close in terms of the trace distance, induced by the trace norm, so $A \approx_{\epsilon} B \Leftrightarrow 1/2|| A- B|| \leq \epsilon$.
If $A$ and $B$ are CPTP maps, then $A$ and $B$ are $\epsilon$-close in terms of the distance induced by the diamond norm on maps.

For an operator $M$ with $\| M\| = Tr |M| = Tr \sqrt{MM^\dagger}$ we denote the trace norm.
The angles are typically denoted $\theta,$ and any arithmetic on angles is assumed to be modulo $2 \pi,$ so in particular for any two objects (operators, states, maps) $O(\theta)$ and $O(\theta + 2\pi)$ are identical.

\subsection{Blind quantum computing and local security criteria}

In this work we will consider both locally secure, and composably secure definitions pertaining to delegated quantum computing. Here, for the most part, we adhere to the notation introduced in \cite{DFPR14}.

\subsubsection{Delegated quantum computing}
In general, delegated quantum computing pertains to any two party protocol between Alice (client) and Bob (server),
where Alice evaluates some (quantum) computation using the help of Bob. In general, any such process is captured by the following definition, which is adapted from \cite{DFPR14}.

\DE\label{DQCdef}
A two-party DQC protocol where Alice's system is denoted $A$ and Bob's system $B$, is modeled as a sequence of
CPTP maps $\{\mathcal{E}_i : \lo{AC} \to \lo{AC} \}_{i=1}^N$ and $\{\mathcal{F}_i :
\lo{CB} \to \lo{CB} \}_{i=1}^{N-1}$, which Alice and Bob apply
sequentially to their respective systems and the communication channel
$C$ (which is initially in the state $\ket{0}$).
The initial state of Alice's and Bob's system we will denote $\psi_{AB}$ and the final state, after the last maps of both players have been implemented with
$\rho_{AB} = DQC(\psi_{AB})$.
\ED
 For example, in the
first round Alice applies $\mathcal{E}_1$ to the joint system $AC$, and sends
$C$ to Bob, who applies $\mathcal{F}_1$ to $CB$, and returns $C$ to
Alice. Then she applies $\mathcal{E}_2$, etc. In the last round Alice applies
$\mathcal{E}_N$. 

The minimal requirement for such a protocol to be useful is that it is correct.
In the local security setting (as opposed to a composable security setting), correctness is a statement about the joint state of Alice and Bob $\rho_{AB}$:
\DE
A DQC protocol (with initial state $\psi_{AB}$), defined by maps $\{\mathcal{E}_i\}_{i=1}^{N}$ and $\{\mathcal{F}_i\}_{i=1}^{N-1}$, where the initial state of Alice's system $\psi_A = Tr_B ( \psi_{AB})$ contains Alice's desired input and the description of her computation is \emph{correct} if when both parties play honestly (\emph{i.e.} apply the maps $\{\mathcal{E}_i\}_{i=1}^{N}$ and $\{\mathcal{F}_i\}_{i=1}^{N-1}$ sequentially) the state of Alice's system at the end of the protocol $\rho_{A} = Tr_{B} (\rho_{AB})$ contains the state generated by applying the computation described in $\psi_A$ to the initial state also given in $\psi_A$.
\ED

Of particular interest for this work, are protocols which also guarantee blindness, intuitively capturing the idea that the server learns nothing about the computation Alice evaluated. 
In the local setting, the relevant definition is given as follows \cite{DFPR14}:
\DE
A DQC protocol provides local $\eps$-blindness, if, for every 
adversarial behavior $\{\mathcal{F}_i\}_i$,  there exists a CPTP map $\mathcal{F}:\lo{B} \to \lo{B}$, such that for all initial states $\psi_{AB}$ (which, along with the defined adversarial behaviour, induces the final state $\rho_{AB}$)
 it holds that 
\begin{equation} \label{eq:sa.b} \tr_A (\rho_{AB}) \approx_\eps \mathcal{F} (Tr_{A} (\psi_{AB})).
  \end{equation} 
\ED
The definition above captures the idea that, if a protocol is blind, then whatever the state of Bob's register is at the end of the protocol, given some initial state, the final state of Bob's register is such that Bob could have generated it locally, without ever engaging with Alice, from the initial state alone. In all the definitions we assume that any purification of the initial states are contained in the register of Bob.

This definition is, when applied specifically to UBQC, and in the case of $\epsilon=0$ equivalent to the original definition of blindness in \cite{BFK09}.

\subsubsection{Universal blind quantum computing}
Universal blind quantum computing (UBQC) \cite{BFK09} is a two-party quantum protocol, founded on the framework of measurement-based quantum computation \cite{RB01}. In it, Alice capable only of producing single qubit states in the state $\ket{+_\theta} = 1/\sqrt{2} (\ket{0}+ e^{i \theta} \ket{1}),$ for the 8 angles $\{ \theta = l \pi/4\}_{l=0}^7$.
We will often refer to this set as ``the 8 state set''.

In general, blind quantum computing protocols come in 4 flavours, depending on whether the input (output) is classical or quantum. In this work, we  focus on the classical input- classical output versions, but the protocols provided are given for the general quantum input-quantum output setting.

In UBQC, Alice initially sends $M$ qubits, randomly chosen in the set of 8 states to Bob, and records their state.
Honest Bob entangles the qubits in the so-called brickwork state (see \cite{BFK09}) - the resulting state is universal for measurement based quantum computing. From that point on the UBQC protocol requires classical communication (specifying the measurement angles applied by Bob, and the corresponding measurement outcomes)
only, barring perhaps the quantum output.

The protocol is given in Protocol \ref{UBQC}, and has been proven to be perfectly correct and perfectly blind in both local settings \cite{BFK09}, and in the full composable sense \cite{DFPR14}, which we elaborate on next.

\begin{algorithm}[htbp]

\caption[UBQC protocol]{Universal Blind Quantum Computation}
\label{UBQC}
\begin{flushleft}
\textbf{Alice's input:}
\vspace{-7pt}
\end{flushleft}
\begin{itemize}
 \setlength{\itemsep}{-1pt}
 \item An $n-$qubit unitary map $U$, represented as a sequence of measurement angles  $\{\phi_{x,y}\}$ of a one-way quantum computation over a brickwork state of the size
 $n \times (m+1)$, along with the $X$ and $Z$ dependency sets 
 $D_{x,y}, D_{x,y}^{'}$, respectively.
\item An $n$-qubit input state $\rho_{in}$
\end{itemize}
\begin{flushleft}
\textbf{Alice's output (for an honest Bob):}
\vspace{-7pt}
\end{flushleft}
\begin{itemize}
 \setlength{\itemsep}{-1pt}
 \item The $n-$qubit quantum state $\rho_{out} = U \rho_{in} U^\dagger$
\end{itemize}
\begin{flushleft}
\textbf{The protocol}
\vspace{-7pt}
\end{flushleft}
\begin{enumerate}
 \setlength{\itemsep}{-1pt}
\item  \label{step:client-prep} \textbf{State preparation} 
\vspace{-7pt}
\begin{enumerate}
 \setlength{\itemsep}{-1pt}

\item \label{step:otp} For each $x \in [n]$, Alice applies
  $X^{i_{x}} Z_{\theta_{x,0}}$ to the $\ith{x}$ qubit of the input
  $\rho_{in}$, where the binary values $i_{x}$ and the angles
  $\theta_{x,0} \in \{k \pi / 4\}_{k=0}^{7}$ are chosen uniformly at
  random for each $x$.  This is
  equivalent to encrypting the input with a quantum one-time pad. The
  result is sent to Bob.
\item If $i_x = 1$, Alice updates the measurement angles $\phi_{x,0}$
  and $\phi_{x,1}$ to compensate for the introduced bit flip, see \cite{BFK09} for details.
\item \label{step:theta} For each column $y \in [m-1]$, and each row
  $x \in [n]$, Alice prepares the state $ \ket{+_{\theta_{x,y}}}:=
  \frac{1}{\sqrt{2}}(\ket{0} + e^{i\theta_{x,y}}\ket{1})$, where the
  defining angle $\theta_{x,y}\in \{k \pi / 4\}_{k=0}^{7}$ is chosen
  uniformly at random, and sends the qubits to Bob.
\item Bob creates $n$ qubits in the $\ket{+}$ state, which are used
  as the final output layer, and entangles the qubits received from
  Alice and this final layer by applying $\ctrlZ$ operators
  between the pairs of qubits specified by the pattern of the brickwork
  state $\mathcal{G}_{n \times (m+1)}$.
\end{enumerate}
%\setcounter{saveenum}{\value{enumi}}
%\end{enumerate}
%\end{algorithm}
%\setcounter{algorithm}{0}
%\begin{algorithm}
%\caption{Universal Blind Quantum Computation - Continued}
%
%\begin{enumerate}
%\setcounter{enumi}{\value{saveenum}}
\item  \label{step:computation-interaction}
 \textbf{Interaction and measurement}
 \vspace{-3pt}

 For $y = 0, \ldots , m-1$, repeat\\
 \hspace*{\parindent}\hspace*{\parindent}  For $x = 1, \ldots, n$, repeat
\begin{enumerate}
 \setlength{\itemsep}{-1pt}
\item Alice computes the updated measurement angle $\phi'_{x,y},$ which depends on previous measurement outcomes Bob reported, and random choices ($r_{x,y}$ below) of Alice, see \cite{BFK09} for details.
\item  \label{step:delta}
Alice chooses a binary digit $r_{x,y} \in \{0,1\}$ uniformly at random, and computes $\delta_{x,y} = \phi'_{x,y}  + \theta_{x,y} + \pi r_{x,y}$.
\item Alice  transmits $\delta_{x,y}$ to Bob, who performs a measurement in the basis $\{ \ket{+_{\delta_{x,y}}},
\ket{-_{\delta_{x,y}}} \}$.
\item Bob transmits the result $s_{x,y} \in \{0,1\}$ to Alice.
\item If $r_{x,y} = 1$, Alice flips $s_{x,y}$;
otherwise she does nothing.
\end{enumerate}
 \setlength{\itemsep}{-1pt}
 \item  \label{OutputCorrection} \textbf{Output Correction}
 \vspace{-7pt}
\begin{enumerate}
 \setlength{\itemsep}{-1pt}
\item Bob sends to Alice all qubits in the last (output) layer.
\item Alice performs the final Pauli corrections $ \{Z^{s_{x,m}^{Z} }X^{s_{x,m}^{X}} \}_{x=1}^{n}$ on the received output qubits.
\end{enumerate}

\end{enumerate}

\label{prot:UBQC}
\end{algorithm}

\subsection{Composable security}

Local security definitions typically consider an abstracted protocol, and pose criteria on the form and structure of the (joint) states of the registers of all players involved in a protocol. In some cases it may be clear what the `correct' definition/criterion should be. However, more often, understanding what happens when such a locally secure protocol is embedded in a larger scheme (or simply is not viewed in isolation from the rest of the universe), may be difficult, counterintuitive, and usually requires additional analyses for every such larger scheme calling a protocol as a subroutine.

Composable security has by now some 20 years of history, and three approaches have been typically used, and have also been extended to the quantum setting:  the Universal Composability (UC) framework \cite{Canetti,BM04,Unr10} (originally by Canetti), the Reactive Simulability (RS) framework \cite{PW01,BPW04,BPW07,Unr04} (originally by Backes, Pfitzmann and Waidner), and the Abstract Cryptography (AC) framework \cite{MR11, Mau12, PMMRT15}, developed by Maurer and Renner. In the approaches of UC and RS, the formalism is built bottom-up, first formally defining the smallest building blocks of any network of protocols, at the level of Turing machines and channels. 
In contrast, AC is a top-down approach, which considers very abstract objects first, proves theorems at this level, and then proceeds to instantiate into particular objects (such as quantum maps and quantum channels) when needed.  AC is in this sense more general, and UC/RS can be realized by instantiating the abstract objects of AC as the systems used in UC/RS.
It is, however, not our goal to give a formal description of how these approaches work, but rather to give a basic flavour of what composability is about. 
Following this we will only state the operative definition of security for two-party delegated computing protocols, as this is all we require for this work - in our case, the UC and the AC definitions are equivalent, so we will not have to be specific about the model we use. The minimal description of how AC is instantiated for the case of delegated quantum computing protocols is given in \cite{DFPR14}, and we refer the interested reader to this work, or to previously given references for further details.
 
The main idea behind composable security, is to define security by an appropriate notion of process identity. Specifically, a security definition consists of a full characterization of a (quantum) map with memory, which may have many distinct `input/output ports'  (which, e.g. partition the Hilbert space of the input systems, and typically assign them to individual players). The characterization is given in terms of its input-output behavior only, and we refer to such maps as \emph{ideal functionalities} or \emph{resources}. 

However, most interesting resources are non-trivial, and do not correspond to (physical) resources we typically have at hand.
In this case the basic idea behind composability is to use other resources (which we do have) to construct the resource we wish to have.
The prescription of how available resources are combined to realize another resource is called a protocol.
Thus protocols turn resources into other resources.

Composability frameworks then provide a method for establishing a similarity/identity between processes, specifically between the overall action realized by using some protocol (which uses some simple resources), and the action of the ideal functionality we wish to have.

The main idea behind composability frameworks is that such a relation can be chosen as to preserve arbitrary nestings, while capturing any notion of secutiy.
Assume that some protocol $\pi$, which uses some set of, perhaps very complicated, resources $R$ is process-similar to an ideal functionality $\mathcal{S}.$ 
The latter defines what security means, thus this sentence can be read as ``$\pi$ is a secure protocol, if we use ideal resources $R$''. 
The composability criterion then asks when can we substitute the complicated resources with sub-protocols (which use simpler resources), while maintaining the overall process similarity with $\mathcal{S}.$ 
Another way to phrase this question is ``Is $\pi$ still secure if we use sub-protocols instead of complicated ideal resources $R$?''. 
The main theorems of composability frameworks then guarantee that this will be the case as long as the sub-protocols themselves are similar to the complicated resources, \emph{under the same similarity relation}. This formalizes the intuitive notion of composability: ``The protocol where we use sub-protocols instead of complicated ideal resources $R$ is secure, as long as the sub-protocols are themselves secure''. 
 
 In other words, if one always proves security of some protocol by defining an ideal resource, and establishing such a process similarity, then one can always analyse new protocols by just assuming the (more complex) ideal resources. Then we do not need to worry about the the details of how they themselves are actually implemented. Such abstraction is, generally, not possible if the security definitions are just local.

A simple example of a useful resource (on purpose chosen outside of an obvious cryptographic context) may be a two-party ideal quantum channel. In this case, the corresponding resource (the channel) has two ports/interfaces, which we can label Alice and Bob. On the first port it accepts a quantum state $\rho$ from Alice, and then it outputs the state $\rho$ on the Bob-interface (or port).

In practice however, the channels are never ideal and quite noisy and in fact the fidelity is typically bounded below by some constant.
This problem is usually resolved by error correction: Alice first encodes her state in an error correcting code, sends this through the channel, and the original state is recovered by Bob.
In the language of the AC framework we would say that the error correcting  protocol $\pi =(\pi_A, \pi_B)$, where $\pi_A$ and $\pi_B$ characterize the local actions of Alice and Bob as stipulated by an error correction algorithm, constructs (within $\epsilon$) the ideal channel $\mathcal{R}_{ideal}$ from the resource of the noisy channel $\mathcal{R}_{noisy}$, when everybody behaves correctly.

 The $\epsilon$ arises as the construction is not perfect as even with very thick error correction, in principle, small errors can still occur. What this captures is that the concatenated processes $\pi_A \mathcal{R}_{noisy} \pi_B$ (that is, the using of the noisy channel, prepended by the process realized by Alice's encoding protocol on Alice's interface, and appended by Bob's decoding on Bob's interface side) is close to the ideal process $\mathcal{R}_{ideal}$. 
 Note the protocols $\pi_A, \pi_B$ are also input-output specified processes. In particular, the input to $\pi_A$ is a quantum state, and the output is the encoding of the input. The output of the concatenated process  $\pi_A \mathcal{R}_{noisy}$ is the encoded state which has undergone noise. The input (on the interface connected to the channel) of $\pi_B$ is an encoded state, and the output (on the Bob-interface) is the decoded state.
 
There are many notions of distance one could consider here (and indeed this is done on a very abstract level in AC),  but in concrete settings (and in this paper) one considers ``the distinguishing advantage''. That is, two resources are $\epsilon-$close, if a third entity, called a distinguisher, having access to all the interfaces of one of the two resources, under optimal strategy, has at most $1/2 +\epsilon$ probability of correctly deciding which resource he is interacting with. When the functionalities are CPTP maps, the distinguishing advantage is given by the diamond norm on the CPTP maps.

What we have discussed thus far is how to express what happens when all players (Alice and Bob) are honest -- that is, what happens if Alice and Bob really run the protocols $\pi_A$ and $\pi_B$. In cryptographic settings, but also in real settings where things are imperfect and can fail, it may be the case that one (or both) of the players are dishonest or simply fail, which must also be covered by the framework.
To formalize such events, every ideal resource usually explicitly separates dishonest and honest usage: each interface of the resource has a so-called ``corruption bit'' $c$ which can be set by a player to $0$ or $1$.
If $c=0$, then the interface allows only honest activities. If $c=1$ then additional functionalities of the resource may be opened up, accessible only to a dishonest player, and these are again fully specified.

To illustrate this setting we will now move to a slightly more complicated example, central to this paper - the functionality of ideal blind quantum computing denoted $\mathcal{S}_B.$ 
The resource $\mathcal{S}_B$ is also a two-player functionality, which takes an input from Alice (a specification of a computation, and perhaps a quantum input $\psi_A$). It was first formally defined in \cite{DFPR14} as follows:
\DE
The ideal DQC resource $\mathcal{S}_B,$ which provides both correctness
and blindness, takes an input $\psi_A, U$ at Alice's interface.
 Bob's filtered interface has a
control bit $c$, set by default to $0$, which a dishonest Bob can flip to activate
the other functionalities. 
If Bob sets $c=0,$ $\mathcal{S}_B$  takes no input from Bob, and produces no output on Bob's side. It produces the honest output $U(\psi_A)$ at Alice's interface. This corresponds to the honest play by Bob.
If $c=1,$ Bob's interface outputs the
allowed leak, here the size of the computation,
 and accepts two
further inputs from Bob: a state $\psi_B$ and a description of a completely-positive trace-preserving map
$\mathcal{E}$. The resource then outputs
 $\mathcal{E}\left(\psi_{AB}\right)$, at the interface of Alice.
\ED
%This is illustrated in Fig. \ref{figure} adapted from \cite{DFPR14}. 
In the above definition, we do not assume that the input state of Alice $\psi_A$ is in tensor product with the system held by Bob -- if it is, the output given to Alice, in the dishonest case, could be written as $ \mathcal{E}\left(\psi_{A}\otimes \psi_B\right).$
The corrupt functionalities model the fact that in blind-only protocols Bob can always interfere and mess up Alice's computation, as there is no mechanism for the verification of the output. But note that, regardless of the activities of Bob, nothing except the pemitted leak is ever output on his interface - so the ideal functionality above captures blindness.

Now, the UBQC protocol we have described in the previous section realizes the functionality $\mathcal{S}_B$, more specifically, it constructs $\mathcal{S}_B$, from the resources of just quantum channels and classical channels. If Bob is honest, and blindly runs his side of the protocol (deleting all the systems he used in the process afterward) it is clear that the input-output behavior of the real protocol is identical to the input-output behavior of $\mathcal{S}_B$. Formally,
we write
\EQ{
\pi_A R_{channels} \pi_B = \mathcal{S}_B \bot_B,\label{corr}
}
where $\pi=(\pi_A, \pi_B)$ is the UBQC protocol (including Bob's deletion of his register at the end) and $\bot_B$ is the (almost tivial) protocol which just sets the corrupt bit $c$ of $ \mathcal{S}_B$ to zero. The statement above is the correctness statement of the protocol in the AC language, and the exact equality means that no distinguisher can tell apart the process $\pi_A R_{channels} \pi_B$ from $\mathcal{S}_B \bot_B$ by having access to just the defined (Bob's and Alice's) interfaces. Naturally, the strict equality can be substituted with an approximate equality.

However, if Bob is dishonest, this is manifestly not the case: in the case of the real protocol UBQC, Bob's side collects all the systems Alice sends in the protocol, whereas in the ideal case, he only gets the permitted leak. 
These settings are trivial to tell apart.

Here is where the notion of the simulator (in UC) or converter (in AC) comes into play.
The simulator is simply a protocol which, from an outside point of view, makes the ideal protocol look like the real one, which, critically, has access only to the interface of the corrupt player. 

In the two player case we are considering, when only Bob can be dishonest, the simulator $\sigma_B$ is an interface between the Bob's interface of $ \mathcal{S}_B$ and Bob himself - this gives rise to the functionality $ \mathcal{S}_B \sigma_B$.
We will say the protocol $\pi$ securely constructs $\mathcal{S}_B$ from some other resources (classical and quantum channels) if there exists a simulator $\sigma_B$ such that
\EQ{
\pi_A R_{channels}  = \mathcal{S}_B \sigma_B \label{seccc}
}
where again the equality may be substituted by an approximate equality.

Note, in above, we have removed Bob's part of the protocol on the left-hand side of the expression, which designates that we no longer specify what Bob may do. 
In the context of UBQC, the left hand side guarantees that Alice sends all the qubits to Bob, and responds to whatever Bob may send back, but says nothing about what Bob is actually locally doing.
This, implicitly, takes care of the quantifier over all possible activities of Bob which arise in local definitions (parts of security statements stating that ``for all activities of Bob'' some criterion holds).
Vitally, the simulator $\sigma_B$ only communicates with Bob's side of the interface, so in particular, it has no direct access to Alice's input, aside from what the ideal functionality may provide. This also means that whatever information about Alice's input Bob could have accessed by being malicious in the real protocol, he could equally access in the ideal protocol, by simply running the simulator $\sigma_B,$ internally.  Thus, intuitively speaking, the real protocol ``is at least as secure'' as the ideal functionality, as a malicious Bob may choose to run the simulator, but is not limited to it. 
This constitutes the asymmetric cryptographic ``process identity'' relation in AC we mentioned earlier is defined as follows: a real protocol is process-similar to an ideal protocol if 1) the correctness expression (Eq. (\ref{corr})) holds, and if there exists a simulator, which makes the ideal protocol look like the real one, where we make no assumptions on what Bob does  (Eq. (\ref{seccc})).
The joint statement that both correctness and security are achieved using protocol $\pi$ we write as
\EQ{
R  \stackrel{\pi, \epsilon} \longrightarrow S,
}
which is read ``the protocol $\pi$ constructs the resource $S$ from the resource $R$, within error $\epsilon$'', and stands as an abbreviation for the formal statements:
\EQ{
\pi_A R \pi_B \approx_{\epsilon} \mathcal{S} \bot_B\ \textup{(correctness)}\label{ccor}\\
\exists \sigma_B,\ s.t.\ \pi_A R \approx_{\epsilon} \mathcal{S} \sigma_B\ \textup{(security)} \label{secc}.
}
The generalization to more complex protocols with multiple players are natural, and we demand the existance of simulators for all players which may be dishonest, and we refer the reader to \cite{PMMRT15} for more details on the general framework which shows how AC is instatntiated in the full quantum setting. We will often abreviate the notation $R  \stackrel{\pi, \epsilon} \longrightarrow S$ with \EQ{R \longrightarrow S,} where in the latter we mean that there exists a protocol realising the construction, where the actual protocol and the error will be clear from context.

In this paper we will only be considering two-player protocols, where only Bob is assumed to be malicious.
As the first step we will show that 

\EQ{
MRSP(8), R_{cc}  \stackrel{\pi, \epsilon=0} \longrightarrow \mathcal{S}_{B},
}
where $R_{cc}$ are classical channels, and $MRSP$ functionality is an abstraction, and a generalization, of Alice's capacity to produce and send the eight qubit states as in the UBQC protocol. It is a generalization in the sense that a dishonest Bob can use it to generate a more general class of classical-quantum correlations than what just direct qubit sending would allow. The formal statement above states such a functionality still suffices for secure blind quantum computing.

Then, we will define ever simpler ideal functionalities  $MRSP(4)$ (using 4 states) and $SP(2)$ (using 2 states) and prove
\EQ{
  MRSP(4) \rightarrow MRSP(8)\ \textup{and}\ SP(2) \rightarrow MRSP(4).
}  
By the composition theorems of AC, we will then have $SP(2) \rightarrow \mathcal{S}_{B},$ where $SP(2)$ requires Alice to produce just two non-orthogonal quantum states (we assume classical channels are always available so we sometimes do not specify them as a required resource explicitly). 
All constructions are with zero error, except $SP(2) \rightarrow MRSP(4),$ where the error is exponentially small in the security parameter.

The paradigm for security is always as follows:
we first identify and fully characterise the desired ideal resource. Then we provide an actual protocol, and prove correctness, which just means that if Bob is honest the output is exactly what we want (Eq. (\ref{ccor})). 
Lastly we prove security, which means we explicitly construct a simulator, which makes the ideal protocol indistinguishable from the real protocol (Eq. (\ref{secc})), when we make no assumptions on what the distinguisher does to find out which functionality, the real or the simulated, he is facing.
Since all the protocols we will encounter are two-round, we will be able to reduce the distinguishing advantage measure to the trace distance on the final states of the real and simulated protocols.

\section{Connections between blind computing and complexity theory}
\label{connections}
The interest in the classical limits of blind quantum computing, especially in its verifiable variant, has been long standing. The capacity to perform universal, and verifiable quantum computation would yield an affirmative answer to the long standing question of Gottesman, Vazirani and Aaronson \cite{Vazirani12,Aaronson07} (GAV), which asked whether one can verify the computation run by a quantum server, using only a classical prover -- formally, whether interactive proof systems, where the prover is limited to BPP, contain the class BQP.
Much of the recent interest in blind quantum computing has been motivated from this perspective.

What is less known (but still established in the narrow scientific community), is that the question of whether just blind (and not verifiable) blind quantum computation is possible with a classical client (the CUBQC question), has independent connections to complexity theory.

Note, the two questions are independent, unless it is proven that any (also classical) blind quantum computing protocol can be lifted to a verifiable protocol, or in reverse, that verifiability automatically implies blindness.

To understand what the possible implications of the (non-)existance of a CUBQC protocol may have, we roughly sketch out what such protocols constitute.
We assume a client and a server, connected by a classical channel.
The client is given as input a description of a unitary $U$ (perhaps in the form of its circuit) over $N$ qubits, and the client is to use the server to come up with the measurement outcome of one of the qubits of $U\ket{\bar{0}}$ where $\ket{\bar{0}}$ is some fiducial, say ``all-zero'', state.
Moreover, we can promise that the outcome (which may be probabilistic) is either $0$ with probability above $2/3$ or $1$ with probability above $2/3$ \cite{ABE10}, as this promise problem is $BQP$-complete.

The client engages in an interaction with the server, where the messages of the client may depend on the input and random bits, generated by the protocol. The server responds to the messages of the client round by round. In the case of an honest server, after at most $poly(N)$ rounds of communication (each round having at most $poly(N)$ bits), the client obtains the answer with bounded error. Importantly, the only thing the server learns (independent of its behavior) is the allowed leak, that is, the bound on the size of the computation $poly(N)$.

Some implications are trivial.
First, if $BQP=BPP$ then such a protocol exists: the client computes the outcome herself, and sends gibberish to the server. By considering the contraposition of this example, we obtain the first simple observation:
\LE
If a CUBQC protocol, where the client is restricted to some computational class $\mathcal{C}$ (say BPP), does not exist, then $\mathcal{C} \not= BQP$ (so $BPP \not= BQP$).
\EL
Thus proving the non-existance of the protocol may be non-trivial, as it would immediately separate the classes BPP and BQP, unless additional restrictions are imposed.

Under certain conditions, the opposite result, establishing the existence of a CUBQC protocol, also has non-trivial consequences.
The main results supporting this, have been established in the paper \cite{AFK}, where the authors study the problem of hiding information from an oracle.
They consider two-party protocols which are called (efficient) Generalized Encryption Schemes, which assume deterministic encryption and decryption
algorithms $E$ and $D$, a randomized key-sampling algorithm $k( \cdot)$. 
\DE \textup{(Generalized Encryption Scheme \cite{AFK})}
 A generalized encryption scheme (GES) for a function $f$, given input $x\in Dom(f)$, is a two-party protocol
with the following properties:
\begin{enumerate}
\item There are m rounds of communication. Denote A's $i^{th}$ message by $a_i$ and
B's $i^{th}$ message by $b_i$.
\item On cleartext input $x$, A computes an encryption key $k = k(x)$ before she
initiates communication with B. This one key is used by A in each round and in
decryption.
\item The encryption algorithm $E$ takes inputs of the form $(x, k, \bar{b})$, where
$x\in Dom(f)$, $k\in \mathcal{K}$, and $\bar{b}$ is a list of elements of the set $\mathcal{B}$ (set of Bob's messages), and produces outputs in $\mathcal{A}$ (set of Alice's messages). In
round $i$ of the protocol, $A$ computes $a_i = E(x, k, \bar{b})$ and sends it to B; the list $\bar{b}$ must
be of length $i - 1$ and consist of B's responses $b_1, \ldots, b_{i-1}$ to A's previous queries.
Thus, in round 1, the list $\bar{b}$ is empty, and the query $a$, is a function of $x$ and $k$.

\item $B$ draws his responses $b_1, ldots, b_m$ from any distribution over $\mathcal{B}$ which satisfies
the property below.

\item The decryption algorithm $D$ takes inputs of the form $(x, k, \bar{b})$, where
$\bar{b}$ is a list of $m$ elements of $\mathcal{B}$. If $k$ and $\bar{b}$ are the results of an execution of
the protocol, then $D(x, k, \bar{b}) =f(x)$ with probability at least$ l/2 + 1/|x|^{c}$, for some
constant $c.$
\item A GES is efficient if $E$, $D$ terminate in polynomial time, and $k(\cdot)$ terminates in expected polynomial time,
we can check in polynomial time whether a particular key $k$ is valid for a cleartext
instance $x$, and the sizes of $k(x),$ $m$ and all the messages are polynomial in the input size $|x|.$ 
\end{enumerate}
\ED

The main result of the paper \cite{AFK} is the following:
\TH
 If the boolean satisfiability problem $SAT$ allows a GES (that is, in the terminology of the definition, $x$ is a boolean formula and $f(x)$ is a statement about its satisfiability), while leaking just a function of the size of the input, then the polynomial hierarchy collapses to the third level.
\HT

The setting of GES is not identical to the setting of CUBQC. In CUBQC, we assume that the server is a quantum machine, capable of computing the class $BQP$. In the setting of GES, however, there is no assumption on the power of the server. Note that the definitions of GES specify the characteristics of Alice's part of the protocol.
Specifically, we may ask the question of existence of GES for SAT, where the server is not omnipotent, but bounded to polynomially-sized quantum computation.
If we require information-theoretic security, and no leak aside from the size, then any GES where the server is bounded to $BQP$ is also a valid GES where the server is not bounded.
This implies that 
\COR
If the boolean satisfiability problem $SAT$ allows a GES (that is, in the terminology of the definition, $x$ is a boolean formula and $f(x)$ is a statement about its satisfiability) where the server is bounded to BQP computations, while leaking just a function of the size of the input against an unbounded server, then the polynomial hierarchy collapses to the third level.
\ROC

Another distinction between CUBQC protocols and GES, is that GES can be specialized for a given function $f$, where CUBQC is intended to be universal.
For this, we can define UGES, with the same specification as GES, where we provide $f$ as an input as well - in other words, in the definition of GES, we substitute $x$ with $(f(\cdot),x)$.

Suppose now that there exists a CUBQC, which is information-theoretically secure, leaks just the size of the input, and the actions of Alice are limited to the specification of UGES, and is capable of securely computing any $BQP$ function.
Then there exists a GES with a BQP-bounded server, capable of computing any $BQP$ function with the same leak, which is also information theoretically secure (this is just an instatiation of UGES where $f$ is fixed). Moreover, since it is information-theoretically secure, it is secure when the server is unbounded.
This means that here exists a GES capable of computing any $BQP$ function.

Thus we have the following observation:
\COR
If there exists an information-theoretically secure CUBQC (where Alice is restricted to the behavior as in UGES), capable of computing any $BQP$ function, while leaking just the size of computation, then 
\EQ{
\textup{If}\ BQP \supseteq NP, \textup{then}\  PH\ \textup{collapses\ to\ the\ $3^{rd}$\ level},
}
\ROC
where $PH$ stands for the polynomial hierarchy,
which to our knowledge would be a new result answering to the affirmative the question posed in \cite{Aa10}.
This alone suggests that secure blind protocols for the CUBQC setting may be hard to find, but we note that the result above assumes perfect security, and a slightly restricted behavior on the side of Alice (as in GES).
However, the restrictions on Alice are slight: note that the function $k$ can serve as a source of polynomial-sized randomness, which can effectively turn the algorithms $E$ and $D$ to randomized algorithms.
Moreover, as stated, the protocol seems to assume that $E$ does not take the queries of Alice as inputs. But, the specification does not preclude that either $E$ re-generates all previous queries, or that the queries of Alice are contained in Bob's responses as well.
Thus the only important source of restrictions comes in that the function $k(\cdot)$ is in the class $ZPP$ which could be be generalized to $BPP,$ and it is not clear whether the main theorem would still hold.

The results of this section motivate the general question of whether a classical client UBQC protocol is possible. In this work, we will tackle the more humble question of realizing UBQC using quantum devices on the side of Alice, but where the devices are further restricted than previously proposed.

\section{The CQ correlations in UBQC}
\label{correlations}

In the classical input/output UBQC, the only quantum phase of the protocol is Alice's preparation - Alice has to prepare and send a bunch of qubits in the state $\ket{+_{\theta}}$ where the phase angle $\theta \in \{k\pi / 4\}_{k=0}^{7} = \Theta$ is chosen uniformly at random.
Similarly, for a verifiable variant of UBQC the set of states Alice has to be able to produce is just augmented by the two Pauli-Z eigenstates.
Alice's capacity to produce such states is a statement about the required classical-quantum correlations of the joint Alice - Bob system which are sufficient for (verifiable) UBQC.
This we can formalize by introducing a third party - Charlie the state preparator.
In the first step of the UBQC protocol, Charlie prepares the following initial state of Alice's and Bob's system:
\EQ{
\sigma_{AB} = \bigotimes_{i=1}^{N} \left( \sum\limits_{\theta_i \in \Theta}  1/8 \dm{\theta_i} \otimes \dm{+_{\theta_i}}\right) 
} 
The states $\{\ket{\theta} \}_{\theta \in \Theta}$ span an 8-dimensional subspace of Alice's register, and signify classical information, \emph{i.e.} $\langle \theta' \ket{\theta} = 0 \ \iff\ \theta' \neq \theta,$ and $N$ is the size of Alice's desired computation.

From this point, the protocol would follow the steps of original UBQC, and clearly the protocol with Charlie the state preparator would be equally (locally) secure. 
Since the original UBQC protocol is blind for every action of Bob, it follows that if Charlie himself cheats, and prepares the state
\EQ{
\sigma_{AB}^{\{ \mathcal{E} \}_i} = \bigotimes_{i=1}^{N} \left( \sum\limits_{\theta_i \in \Theta} 1/8  \dm{\theta_i} \otimes \mathcal{E}_i (\dm{+_{\theta_i}})\right) \label{suff}
} 
where the CPTP maps $\mathcal{E}_i$ were chosen by Bob, the resulting protocol would still be blind (albeit not necessarily \emph{correct}) \footnote{In the expression above, we assume Bob/Charlie's deviation is separable, however blindness would equally hold in the case of a coherent deviation, where the realized joint state is of the form
\EQ{
\sigma_{AB}^{ \mathcal{E} } = (\Id_A \otimes \mathcal{E}_B)\bigotimes_{i=1}^{N} \left( \sum\limits_{\theta_i \in \Theta}  \underbrace{\dm{\theta_i}}_{A} \otimes  \underbrace{\dm{+_{\theta_i}}}_{B}\right).
} 
In this note we investigate the characteristics of the simplest apparatus Alice may use and still achieve blindness, so we focus on properties of individual states. We note, however, that the reduced density matrix obtained by tracing out all but one individual system in the expression above again generates individual states with the correlations given by Eq. (\ref{suff2}).
}.
Focusing on individual states, the sufficient correlations then attain the form
\EQ{
\sigma_{AB - single}^{\mathcal{E}} =  \sum\limits_{\theta \in \Theta}  1/8 \dm{\theta} \otimes \mathcal{E}(\dm{+_{\theta}}) \label{suff2}
} 
Similarly, for verifiable UBQC \cite{FK12}, Charlie would generate such correlations with the only difference that a certain number\footnote{This number depends on the desired verifiability levels of verifiable UBQC, and it is proportional to the number of testing qubits, called trap qubits, which are used in verifiable UBQC to monitor the acitivities of Bob.} of subsystems are characterized with the correlation
\EQ{
\sum\limits_{b=0}^{1}1/2 \dm{Z_b} \otimes \dm{b},
} such that $\bra{Z_b}\theta \rangle = 0, \forall \theta, b$ and  $\bra{Z_0} Z_1\rangle =0$. 
where blindness is again ensured as long the deviation from the ideal correlations is characterized by a CPTP map acting on Bob's register only (as it corresponds to a possible deviation by Bob at the very first step of the protocol).
The sufficiency of establishing the correlations of the type in Eq. (\ref{suff}) has already been used in \cite{DKL12} to provide a coherent state UBQC protocol through a coherent state remote blind qubit state preparation sub-protocol.

The correlations of the type given in Eq. (\ref{suff2}), perhaps with the additional states required for the verifiable protocol as explained above, suffice for obtaining a verifiable protocol of the type given in \cite{FK12}. We will call such correlations \emph{correlations for verifiable UBQC} (sometimes \emph{strong correlations}), as these types of correlations are sufficient for both blindness and verifiablity. 

\subsection{Necessary individual correlations for blindness}
While the individual correlations given in  Eq. (\ref{suff}) are \emph{sufficient} for blindness of UBQC, they are not \emph{necessary}.
The following Theorem gives the characterization of necessary and sufficient individual correlations for blind UBQC:
\TH \label{necc-ind} A UBQC protocol, with classical input with computation of size $N$ where the preparation stage, in which Alice emits her pre-rotated qubits, is replaced by initializing a part of the joint system of Alice and Bob to $N$ states $\sigma_{AB}^{i}$ is blind if and only if the states $\sigma_{AB}^{i}$ are of the following form:
\EQ{
\sigma_{AB}^{i} =  \dfrac{1}{|\Theta| }\sum\limits_{\theta_i \in \Theta}  \dm{\theta_i} \otimes \rho_i^{\theta_i}, \label{necc1}
} 
where it holds that
\begin{enumerate}
\item $\rho^{\theta}$ is a normalized quantum state (trace one positive operator), for all $\theta$, and
\item $\rho^{\theta} + \rho^{\theta+\pi} = \rho^{\theta'} + \rho^{\theta'+\pi}$ for all $\theta,\theta'$,
\item $|\Theta|$ is the size of the set $\Theta$, typically 8.
\end{enumerate}
In other words,
for all $\theta$, $\rho^{\theta} + \rho^{\theta+\pi} = 2 \eta_i$ where $\eta_i$ is some fixed quantum state.
The correlations of the type given in Eq. (\ref{necc1}) we will call  \emph{weak} correlations, or \emph{correlations for blindness}.
\HT
\begin{proof}
\noindent\textbf{Sufficiency:}\\

Following the steps proof of stand alone blindness of UBQC given in \cite{Dunjko12}, for all deviations Bob may implement, the state of Bob's register at the end of the computation can be written as:
\EQ{
\mathcal{N} \sum_{\ora{r}} \sum_{\ora{\theta}}  \sum_{\ora{b}} \mathcal{E}^{\ora{b}}  \left(  \bigotimes_{i=1}^{N} \left( \rho^{\theta}_i  \otimes   
 \bb{\delta_i(\ora{b}, \ora{r}, \theta_i,\phi_i)} \right) \otimes \psi_B 
 \right)
}
where
\begin{itemize}
\item $\ora{b}, \ora{r}, \ora{\theta}$ and $\ora{\phi}$ denote the vectors of length $N$ comprising Bob's classical responses throughout the protocol, the random bits $r$ chosen by Alice, the random angles $\theta$ coming from the initial pre-set correlations, and Alice's computational angles, 
\item $\psi_B$ denotes the initial state of Bob's system 
\item $\mathcal{E}^{\ora{b}} $ denote completely positive trace non-increasing maps Bob implements depending on the declared measurement outcomes he sends to Alice 
\item for a label $X$ we abbreviate $\dm{X}$ with $\bb{X}$, so with 
$ \bb{\delta_i(\ora{b}, \ora{r}, \theta_i,\phi_i)}$ we denote the classical measurement angles Alice sends to Bob throughout the protocol, which may depend on all parameters.
\item $\mathcal{N}$ denotes the normalization factor $2^{-(4 N)}$ corresponding to the number of Alice's possible choices of $r$ and $\theta$ parameters.
\end{itemize}

Since CP maps are linear, the expression above rewrites as:
\EQ{
\mathcal{N}    \sum_{\ora{b}} \mathcal{E}^{\ora{b}}  \left( \sum_{\ora{r},\ora{\theta}} \left( \bigotimes_{i=1}^{N} \left( \rho^{\theta}_i  \otimes   
 \bb{\delta_i(\ora{b}, \ora{r}, \theta_i,\phi_i)} \right) \right) \otimes \psi_B 
 \right), \label{expr:rewritt}
}
and to prove the sufficiency of the theorem, it will suffice to show that 
\EQ{
 \sum_{\ora{r},\ora{\theta}} \left( \bigotimes_{i=1}^{N} \left( \rho^{\theta}_i  \otimes   
 \bb{\delta_i(\ora{b}, \ora{r}, \theta_i,\phi_i)} \right) \right)
}
is an operator independent from the computational angles.

To show this, we will use the following lemma:
\LE \label{lem:eta} Let $\delta(\theta, r) = \phi + \theta + r \pi \ (\textup{mod}\ 2 \pi)$, for a fixed angle $\phi$, be a measurement angle as appearing in UBQC, and $\rho^{\theta}$ be a state such that $\rho^{\theta} + \rho^{\theta+\pi} = 2 \eta $ for all $\theta$ and some fixed state $\eta$ (independent of $\theta$). Then
\EQ{
 \sum\limits_{\theta, r} \rho^{\theta} \otimes \bb{\delta(\theta, r)} = 2 \eta \otimes \Id.
}
\EL 
\begin{proof}
We prove this lemma by reshuffling the elements of the sum. Note that we have, for all $\phi$, that $\delta(\theta, 0) =  \delta(\theta+\pi, 1)$, so the sum above rewrites as
\EQ{
\sum\limits_{\theta = 0}^{3\pi/4} (\rho^{\theta} + \rho^{\theta+\pi}) \otimes \left(\bb{\delta(\theta, 0)} + \bb{\delta(\theta,1)} \right) = 2 \eta \otimes \sum\limits_{\theta} \left(\bb{\delta(\theta, 0)}\right) = 2 \eta \otimes \Id
}
\end{proof}

Now we return to the evaluation of the expression
\EQ{
 \sum_{\ora{r},\ora{\theta}} \left( \bigotimes_{i=1}^{N} \left( \rho^{\theta}_i  \otimes   
 \bb{\delta_i(\ora{b}, \ora{r}, \theta_i,\phi_i)} \right) \right)
}

Note that, for a fixed sequence of Bob's responses $\ora{b}$ we can, by the definition of the angles $\delta$, introduce a substitution of variables by introducing the variables 
$r_i' = (r_i + b_i)\ \textup{mod} \ 2$. This corresponds to the fact that, in the UBQC protocol, Alice internally flips her hidden $r$ parameters depending on the response of Bob.
In these new variables, the angles $\delta$ only depend on $r'$ which are still distributed uniformly at random.
The expression above then becomes  
\EQ{
 \sum_{\ora{r'},\ora{\theta}} \left( \bigotimes_{i=1}^{N} \left( \rho^{\theta}_i  \otimes   
 \bb{\delta_i( \ora{r'}, \theta_i,\phi_i)} \right) \right)
}
Note that the last $r'$ variable $r'_N$ only appears in the angle $\delta_N$, and the variable $\theta_i$ only appears in the corresponding angle $\delta_i$.
So we can break up the sum above as:
\EQ{
 \sum_{r'_1,\ldots r'_{N-1},\theta_1, \ldots \theta_{N-1}}  \bigotimes_{i=1}^{N-1} \left( \rho^{\theta}_i  \otimes   
 \bb{\delta_i( \ora{r'}, \theta_i,\phi_i)} \right)  \otimes \left( \sum\limits_{r'_N, \theta_N}\rho^{\theta}_N  \otimes   
  \bb{\delta_N( \ora{r'}, \theta_N,\phi_N)} \right)
}
However, by Lemma \ref{lem:eta}, and the assumptions of the Theorem we have that
\EQ{
\sum\limits_{r'_N, \theta_N}\rho^{\theta}_N  \otimes   
  \bb{\delta_N( \ora{r'}, \theta_N,\phi_N)} = 2 \eta_N \otimes \Id.
}
Since, for every $i$ the modified computational angle $\phi_i'$ appearing in $\delta_i$ only depends on $r'_j$ parameters where $j<i,$ the process of eliminating the last remaining angle $\delta$, as shown for $\delta_N$ above, continues inductively until we obtain
\EQ{
 \sum_{\ora{r'},\ora{\theta}} \left( \bigotimes_{i=1}^{N} \left( \rho^{\theta}_i  \otimes   
 \bb{\delta_i( \ora{r'}, \theta_i,\phi_i)} \right) \right) = \bigotimes_{i=1}^{N} \left( 2 \eta_i \otimes \Id  \right).
}
by plugging this into expression in (\ref{expr:rewritt}) we get that the state of Bob's register is: 

\EQ{
\mathcal{N}    \sum_{\ora{b}} \mathcal{E}^{\ora{b}}  \left(  \bigotimes_{i=1}^{N} \left( 2 \eta_i \otimes \Id  \right)  \otimes \psi_B ,
 \right) =  \mathcal{E}(\psi_B),
}
for some CPTP map $\mathcal{E}$. That is the protocol is stand-alone perfectly blind.

\noindent\textbf{Necessity:}\\
For a protocol to be blind, it should be blind for every size of the computation, in particular for the computation of size $N=1$, and also in the case when, as his first step, Bob deletes all information in his register except the initial correlations.
Then the reduced state of Bob's system is given with:
\EQ{
\sum_\theta \rho^{\theta}.
}
As the next step of the protocol,  Alice sends her first and only measurement angle $\delta(\theta, r, \phi)=\phi + \theta + r \pi \ \textup{mod} \ 2 \pi,$  and Bob's reduced state updates to \EQ{
\sum_{\theta, r} \rho^{\theta} \otimes \bb{\delta(\theta, r, \phi)},
}
up to an irrelevant normalization factor,  where Alice's  secret measurement angle is $\phi$.
In order for the protocol to be stand alone blind (or composably blind, for that matter), the state above must not depend on $\phi$.
The sum above can still be reshuffled, since $\delta(\theta, 0, \phi) = \delta(\theta+\pi, 1, \phi)$, to obtain
\EQ{
\rho_{\phi} = \sum\limits_{\theta = 0}^{3\pi/4} (\rho^{\theta} + \rho^{\theta+\pi}) \otimes \left(\bb{\delta(\theta, 0,\phi)} + \bb{\delta(\theta,1,\phi)} \right).
}
Note that the operators $\bb{\delta(\theta, 0,\phi)} + \bb{\delta(\theta,1,\phi)} $ and $\bb{\delta(\theta, 0,\phi')} + \bb{\delta(\theta,1,\phi')} $ have orthogonal support unless $\phi = \phi'$ or $\phi = \phi' + \pi$, so it can be shown that if $\rho_{\phi} = \rho_{\phi'}$ for all $\phi,\phi'$ then $\rho^{\theta} + \rho^{\theta+\pi} = \rho^{\theta'} + \rho^{\theta'+\pi}$ for all $\theta, \theta'$.
Thus, the criterion of the Theorem is necessary for (perfect) blindness as well.
\end{proof}

We emphasize that the weak correlations defined in Theorem \ref{necc-ind} are strictly weaker than the strong correlations defined in Eq. (\ref{suff2}).
Note first that strong correlations also satisfy the condition for weak correlations as
\EQ{
\mathcal{E}(\dm{+_\theta})+\mathcal{E}(\dm{+_{\theta+\pi}}) = \mathcal{E}(\dm{+_\theta}+\dm{+_{\theta+\pi}}) = \mathcal{E}(\Id), \forall \theta.
}
However, the converse is not true, as for instance the following function $f$:
\EQ{
f(\theta) = \dm{+_{3 \theta}} \label{functionF}
}
defines the correlations
\EQ{
\sum\limits_{\theta} \dm{\theta} \otimes f(\theta) = \sum\limits_{\theta} \dm{\theta} \otimes \dm{+_{3 \theta}}
}
which are weak correlations for UBQC since 
\EQ{
f(\theta) +f(\theta+\pi)  = \dm{+_{3 \theta}} +\dm{+_{3 \theta+ 3\pi}} = \Id.
}
But, these are not strong correlations since $f$ is not a linear map, hence not a quantum channel.
This is the underlying reason why the two-server blind quantum computing protocol in \cite{BFK09} cannot directly be used to achieve verifiable blind quantum computation - the two server setting uses non-communicating servers for the preparation phase, and it is easily proven that the resulting correlations are the weak correlations, for all strategies of the two servers. However, there exist strategies which generate weak correlations which are manifestly not strong, thus the verifiability theorems of the verifiable protocol in FK12 \cite{FK12} cannot be applied. We will come back to the two-server setting again later.
%
%\subsection{Sufficient and necessary individual correlations for approximate $(\epsilon>0)$ blindness}
%It is easy to see, by the union bound, that, if one is interested in approximate $\eps$-blindness then the capacity to establish individual correlations of the form
%\EQ{
%\sigma_{AB}^{i} \approx_{\delta}  \sum_{\theta_{i} \in \Theta}  \dm{\theta_i} \otimes \rho_i^{\theta_{i}},
%} 
%where the properties of the states $\rho_\theta$ are as given in Theorem \ref{necc-ind} are sufficient where $\delta \leq \eps/N$ where $N$ is the computation size.
%
%
%Establishing the necessary criteria for individual correlations for approximate $\eps$-blindness is far more involved. The parameter $\eps$ effectively quantifies the trace distance between the system of Bob obtained by starting from ideal individual correlations and the system obtained by starting from approximate correlations. 
%Without having more information on the particular nature of the deviation between the ideal and approximate individual states, aside from the trace distance of individual systems, the best bound on the trace distance of the collective systems is given by multiplying the largest individual trace distance with the total number of systems (\emph{i.e.} by the union bound).
%Since we are interested in the general properties of UBQC, here we will not investigate the cases in which tighter bounds can be found.
%
%
The results of this section provide simple criteria characterising the properties of the initial correlations between Alice and Bob, which must be guaranteed in order for the overall protocol to be secure in a local sense. If we simply abstract the process which establishes such correlations (rather than the end result) we identify the ideal functionality which can be used by Alice to realize delegated quantum computation. This naturally leads us to the composable security framework.

\section{Ideal functionalities for blind UBQC}
\label{functionalities}
In the previous sections, we have characterized initial correlations and protocols which are sufficient or necessary for stand-alone blindness of UBQC.

Here, we formalize the establishing of these pre-existing correlations by an ideal process -- an ideal functionality -- which will then allow us to study other protocols which can be used instead of the standard state preparation. In particular, we will give protocols which securely realize the ideal preparation functionalities with lower demands on Alice.

\subsection{Basic preparation functionalities}
The most natural functionality which generates the weak correlations comes straightforwardly from their specification:

\DE
The ideal resource called the \emph{random remote blind state preparation for blindness} $(RSP_B)$ has two interfaces A, and B, standing for Alice and Bob.
The resource first selects a $\theta$ chosen uniformly at random. Bob's interface has a filtered functionality comprising a bit $c$ which Bob can pre-set to zero or one, depending on whether he will behave maliciously. If Bob pre-sets $c=0$, the resource outputs the state $\dm{+_{\theta}}$ on Bob's interface.
If Bob pre-sets $c=1$, it awaits the set $\{(\theta, \left[ \rho^{\theta}\right] )\}_{\theta}$ from Bob, where $\left[ \rho^{\theta}\right]$ denotes the classical description of a quantum state, with the property that $\rho^{\theta} + \rho^{\theta+\pi}  =\rho^{\theta'} + \rho^{\theta'+\pi}, \forall \theta,\theta'.$ 
If the states Bob inputs do not satisfy the property above, the ideal functionality ignores the set Bob has input and awaits a new valid set.
Once the set is received, the functionality outputs $\rho^{\theta}$ at Bob's interface. In both cases, the resource outputs the angle $\theta$ at Alice's interface.
\ED

It should be clear that the resource above has been tailored to establish exactly the weak correlations.
However, it will be useful to define another type of a functionality, which is more in the measurement based spirit. Here, instead of receiving a specification of states, the functionality receives a specification of measurements by Bob, and a system, which can be used, using entanglement, to steer Bob's state.

\DE
The ideal resource called  measurement-based remote blind state preparation for blindness $(MRSP_B)$ has two interfaces A, and B, standing for Alice and Bob.
The resource first selects a $\theta$ chosen uniformly at random. Bob's interface has a filtered functionality comprising a bit $c$ which Bob can pre-set to zero or one, depending on whether he will behave maliciously. If Bob pre-sets $c=0$, the resource outputs the state $\dm{+_{\theta}}$ on Bob's interface.
If Bob pre-sets $c=1$, it awaits the set of descriptions of $8$ positive operators $\{ \Pi_{\theta} \},$ such that for all $\theta,$ $ \Pi_{\theta} +  \Pi_{\theta+\pi}  = \mathbbmss{1}.$ Thus a pair of such operators differing by $\pi$ constiutes a complete measurement.  
Additionally, it accepts a system $B_{in}$ in some quantum state, of the same dimension as the operators $\Pi_\theta$.
Once these have been input, the functionality chooses a random angle $\theta,$ and performs the complete measurement $\Pi_{\theta} , \Pi_{\theta+\pi},$ and outputs the outcome ($\theta$ or $\theta+\pi$) on Alice's interface.
If the descriptions of operators (or dimensionality of $B_{in}$) Bob inputs do not satisfy the described properties, the ideal functionality ignores the set Bob has input and awaits a new valid set.
Once the set is received, the functionality outputs $\rho^{\theta}$ at Bob's interface. In both cases, the resource outputs the angle $\theta$ at Alice's interface.
\ED

Both functionalities can naturally be defined for any even number of angles (then denoted in parenthesis explicitly, e.g. $MRSP_B(8)$), which we shall use later.

The latter resource allows for a larger set of options for a dishonest player. This is captured by the fact that we can construct $MRSP_B$ from $RSP_B$.
It is a good warm-up exercise to prove this. The protocol  $\pi = (\pi_A, \pi_B)$ is trivial - Alice does nothing and Bob presses the no-corrupt button (sets $c=0$).
Recall to prove that the trivial protocol indeed constructs $MRSP_B$ from $RSP_B$ we need to show the following identities:
\EQ{
\pi_A RSP_B \pi_B = MRSP_B \bot_B
}
which is the correctness part, and we need to find the protocol which is the simulator $\sigma_B,$ such that

\EQ{
\pi_A RSP_B  = MRSP_B \sigma_B \label{sec}.
}
The correctness follows from the definitions of the resources, as both output the same thing in the honest case.

To find the simulator and prove security, we will first introduce a steering-type lemma we will be using throughout the paper.

\LE\label{steer}
Let $\rho^\theta$ be any set of 8 states, parametrized by the 8 angles $\theta$, such that for all $\theta$
\EQ{
\rho^\theta + \rho^{\theta+\pi} = 2 \eta,
}
where $\eta$ is a density matrix, independent from the angle $\theta$.
Then there exist a set of measurement operators  $\{ \Pi_{\theta} \},$ satisfying $ \Pi_{\theta} +  \Pi_{\theta+\pi}  = \mathbbmss{1},$ and a bipartite pure state $\ket{\eta}_{12}$ such that for every $\theta$, the preforming of the complete measurement characterized by $\{ \Pi_{\theta} , \Pi_{\theta+\pi} \}$ on subsystem $1$, and obtaining the outcome $\theta+b \pi$ (for a bit $b \in \{0,1\}$), leaves the subsystem $2$ in the state $\rho^{\theta+ b\pi}$. The probabilities of the two outcomes are uniform for any $\theta$.
\EL
\begin{proof}
This is a relatively simple steering-type result.
Let
\EQ{
\eta = \sum\limits_{k=1}^{M} \lambda_k \dm{\psi_k} 
} be the spectral decomposition of $\eta$,
where $\lambda_k>0$.
Then we set the state $\ket{\eta}_{AB}$ to
\EQ{
\ket{\eta}_{AB} = \sum\limits_{k=1}^{M} \sqrt{\lambda_{k}} \ket{\psi_{k}}_{A}\ket{\psi_{k}}_{B}. 
}

Next, for every pair $\theta,$ $\theta+\pi$, we define the following operators:
\EQ{
\Pi^\theta  = 1/2 \eta^{-1/2} {\rho^{\theta}}^{\tau}   \eta^{-1/2} \\
\Pi^{\theta+\pi}  = 1/2 \eta^{-1/2} {\rho^{\theta+\pi}}^{\tau}  \eta^{-1/2},
}
where $\eta^{-1/2} = \sum\limits_{k=1}^{M}1/\sqrt{\lambda_k} \dm{\psi_k} $ \footnote{Such measurement operators are intimately related to the so-called ``square-root'' measurements \cite{Eldar00} appearing in optimal measurements literature.}.

First, note that $\Pi^\theta+\Pi^{\theta+\pi}$ is the identity on the subspace where the state $\eta$  has non-zero support.
\EQ{
1/2 \eta^{-1/2} {\rho^{\theta}}^{\tau}   \eta^{-1/2} +  1/2 \eta^{-1/2} {\rho^{\theta+\pi}}^{\tau+\pi}  \eta^{-1/2}  =   1/2 \eta^{-1/2} {2\eta}^{\tau}   \eta^{-1/2} = \sum\limits_{k=1}^{M} \dm{\psi_k} ,
}
where the last equality holds as $\eta = \eta^\tau$ relative to its eigenbasis.
If the support is not over the entire state space, we formally append the measurement operator set to $\{\Pi^{\theta} ,\Pi^{\theta+\pi} , \mathbbmss{1} - \Pi^{\theta} -\Pi^{\theta+\pi} \},$ but the third outcome will never occur.

Next, we claim $\Pi^\theta, \Pi^{\theta+\pi}$ are positive-semidefinite operators. To see this, we will fix the basis to the set $\{\ket{\psi_{k}} \}$. In this basis, the operator is diagonal, and the matrix of the entire operator can be written as 
\EQ{
{\rho^{\theta}}^{\tau} \circ M,
}
where $M$ is a matrix with elements $\bra{\psi_i}M\ket{\psi_j} = 1/\sqrt{\lambda_{i}\lambda_{j}},$ which is just the outer product of the vector $(1/\sqrt{\lambda_i})_i$ with itself, and where $\circ$ denotes the entry-wise matrix product (sometimes called the Shur, or the Hadamard matrix product).  $M$ is obviously a rank 1 matrix, with one positive eigenvalue. Thus it is positive-semidefinite, but so is ${\rho^{\theta}}^{\tau}, $ and since the entry-wise product of two positive-semidefinite operators yields a positive-semidefinite operator, the entire operator is positive-semidefninite.

Finally, we consider the influence the measurement has on the leftover system. Assume we obtain the outcome associated with $\Pi^\theta,$ then the resulting (subnormalized) state is
\EQ{
\sum_{k,k'} \sqrt{\lambda_{k} \lambda_{k'}} Tr(\Pi^\theta \ket{\psi_k} \bra{\psi_{k'}} )   \ket{\psi_k} \bra{\psi_{k'}}.
}
Then we have
\EQ{
Tr(\Pi^\theta \ket{\psi_k} \bra{\psi_{k'}} ) =\bra{\psi_{k'}}  \Pi^\theta \ket{\psi_k}  = 1/2 \bra{\psi_{k'}} \sum\limits_{l=1}^{M}1/\sqrt{\lambda_l} \dm{\psi_l}  {\rho^{\theta}}^{\tau}  \sum\limits_{h=1}^{M}1/\sqrt{\lambda_h} \dm{\psi_h} \ket{\psi_k} = \\
1/2\sqrt{\lambda_{k'} \lambda_{k}} \bra{\psi_{k'}} {{\rho^{\theta}}^\tau}\ket{\psi_k} = 1/2\sqrt{\lambda_{k'} \lambda_{k}} \bra{\psi_{k}} {{\rho^{\theta}}}\ket{\psi_k'}
 }
 Plugging this back in the whole expression, we get
 \EQ{
 \sum_{k,k'} \sqrt{\lambda_{k} \lambda_{k'}} Tr(\Pi^\theta \ket{\psi_k} \bra{\psi_{k'}} ) \ \ket{\psi_k} \bra{\psi_{k'}} = 
 1/2\sum_{k,k'}\bra{\psi_{k}} {{\rho^{\theta}}}\ket{\psi_{k'}} \ket{\psi_k} \bra{\psi_{k'}} = 1/2\rho^\theta,
 }
 where 1/2 is just the probability of that outcome. We get the analogous result for $\theta+\pi$ outcome measurement, so the Lemma holds.

\end{proof}

The lemma above can naturally be generalized to more measurement outcomes (and more states we remotely prepare), but two will suffice in this work.

Now, the simulator $\sigma_B$ is defined as follows. It has exactly the same Bob-interface as the functionality $RSP_B,$ so if the distinguisher presses $c=0$, it presses $c=0$ on $MRSP_B,$ and forwards what the functionality outputs.
If Bob presses $c=1,$ it presses $c=1$ on $MRSP_B$, and it collects the descriptions of states $\{(\theta, \left[ \rho^{\theta}\right] )\}_{\theta}$ from Bob, and checks if the opposite states (in terms of the angles) add up to the same state $2\eta$ (it awaits for a new valid set if not). If so it uses the results of Lemma \ref{steer} to find the corresponding descriptions of the measurement operators $\{ \Pi_{\theta} \},$ (for the input states described by$\{(\theta, \left[ \rho^{\theta}\right] )\}_{\theta}$) and it prepares the purification of the state $\eta$, outputting one part of it to Bob, and inputs the descriptions of the measurement operators, and the other part of the system which held the purification $\ket{eta},$ to the resource $MRSP_B.$ The resource $MRSP_B$ will perform a measurement as specified, and by Lemma \ref{steer} the joint state of Alice and Bob will be exactly the same as the joint state which would have been achieved by the same actions using $\pi_A RSP_B$. 
This finishes the proof.

The inverse claim, that $RSP_B$ can be used to securely construct $MRSP_B$ does not hold: since Bob chooses what system to input to $MRSP_B,$ and since this is a quantum state, there is no way for the simulator to exactly estimate which states it should request from $RSP_B$.

The main reason why we are studying these local preparation functionalities is to substitute the preparation part of the UBQC protocol, where Alice prepares individual qubit states in one of the 8 states uniformly at random, stores the angle and sends them to Bob, with a functionality which has lower requirements on Alice. 
This process can be abstracted as an ideal functionality where Alice receives an angle, and Bob the corresponding qubit state if he is honest.
If he is not, since the quantum channels allows for no corrupt activity, the only thing Bob can do is apply a local CPTP map to the qubit he receives.

This brings us to a strong functionality corresponding to strong correlations from the previous sections:

\DE
The ideal resource called the strong random remote state preparation $(RSP_{S})$ has two interfaces A, and B, standing for Alice and Bob.
The resource first selects an angle $\theta$ (from the set of 8 states) chosen uniformly at random. Bob's interface has a filtered functionality comprising a bit $c$ which Bob can pre-set to zero or one, depending on whether he will behave maliciously. If Bob pre-sets $c=0$, the resource outputs the state $\dm{+_{\theta}}$ on Bob's interface.
If Bob pre-sets $c=1$, it awaits a description of a CPTP map $\mathcal{E}$ from Bob.  
Once the set is received, the functionality outputs $\mathcal{E}(\dm{+_{\theta}})$ at Bob's interface. In both cases, the resource outputs the angle $\theta$ at Alice's interface.
\ED

Two things should be clear. First, the resource above is weaker than $RSP_B$, in that we can use  $RSP_S$ to construct  $RSP_B$ - the protocol is again trivial, and so is the correctness of the construction (as both functionalities have identical honest behaviors).
For security, we need to show that $RSP_S = RSP_B \sigma_B$ (omitting Alice's part of the protocol as it is trivial), and this is easy. The simulator $\sigma_B$ collects the map $\mathcal{E}$ and inputs the descriptions of the states $\{\mathcal{E}(\dm{+_{\theta}})\}_{\theta}$ to $RSP_B$ (they are valid inputs) and forwards to Bob what is output.

Second, if we substitute the act of Alice sending a qubit to Bob in UBQC with $RSP_S$, the security proof does through trivially, as the latter is just a formal specification of the prior. Moreover, this type of a strong preparation protocol also suffices for a verifiable variant of the UBQC protocol \cite{FK12}, as we had clarified earlier. From the analysis of the last section, however, the intuition should be that weaker resource should do as well.

Indeed, $RSP_B$ is a weaker resource than $RSP_S$ (it allows more options for Bob than $RSP_S$) thus it does not follow trivially that the UBQC protocol is secure if we use this weaker resource. Furthermore, since $MRSP_B$ is weaker still than $RSP_B$, again it does not follow that UBQC is secure if $MRSP_B$ is used. Next, we show that by a simple tweak of the original composable security proof of UBQC, we can substitute the ``qubit generation and sending'' with $MRSP_B$ and still have a secure protocol.

To make this formal, we remind the reader that the original statement of security of UBQC is that the UBQC protocol constructs the ideal resource $\mathcal{S}_{B}$ of blind quantum computing, from classical channels, single qubit preparation, and quantum channels. Now we claim that (classical input - classical output) UBQC constructs $\mathcal{S}$ from classical channels, and calls to the $MRSP_B$ resource.

\TH
\label{UBQCprep}
The UBQC protocol where Alice, instead of access to a quantum channel and a random generator of $\ket{+_{\theta}}$ states, has access to the ideal functionality $MRSP_B$ exactly constructs the ideal functionality $\mathcal{S}_B$. In other words, the UBQC protocol is perfectly secure. \HT
\begin{proof}

We begin the proof by reproducing, verbatim, the protocol detailing the actions of the simulator and the ideal resource given in the paper \cite{DFPR14} as Protocol \ref{prot:sim}.

The basic idea behind this construction is that instead of sending $\ket{+_\theta}$ states, and computing the declared angles $\delta$ from the computational angles $\phi'$, Alice could have equivalently send halves of Bell-pairs, and random $\delta$ angles. Then, the angle $\theta$ would be computed in run-time, and Bob's local state would be steered to either $\ket{+_\theta}$ or $\ket{-_\theta}$ by a local measurement. 
Specifically, the projection of one part of the symmetric Bell-pair $1/\sqrt{2}(\ket{00} +\ket{11})$ with the projector $\dm{+_{-\theta}},$ leaves the remaining system in the state $\dm{+_\theta}.$ Note, this is just a special case of Lemma \ref{steer}, as $\dm{+_{-\theta}}$ is the transpose of $\dm{+_{\theta}}$ relative to the computational basis.

The outcome corresponds to the random bit $r$ in the original protocol. Rephrased like this, the ``steering'' variant of the protocol can be split into two parts -- the preparation part (done by the simulator), and the measurement part, left to the ideal resource.

To prove our theorem, it suffices to modify the simulator as follows: instead of preparing Bell pairs, and sending half to Bob, it collects the system from Bob and the description of measurement operators (as specified by the functionality $MSP_B$) -- this modifies step \ref{step1} of the Protocol \ref{prot:sim}.  It still outputs a random $\delta$ angle.

Later, it simply forwards the system, and the operators, for each system, to the ideal functionality -- modifying step \ref{step2} -- and instructing it to apply the corresponding measurement -- visible in the modification of step \ref{step3} of Protocol \ref{prot:sim}.
The modified protocol is given in Protocol \ref{prot:mod}.
By Lemma \ref{steer}, even in this more general case, the state obtained by Bob will be exactly equal as in the real case. \qed

The functionality $MRSP_B$ was designed with the $\mathcal{S}_B$ functionality in mind, as it takes on input a state and CP maps as well.

\begin{algorithm}[htbp]

\caption[UBQC with and simulator]{Protocol with $\mathcal{S}_B$ and simulator}
\label{prot:sim}
\begin{flushleft}
\textbf{The simulator}
\vspace{-7pt}
\end{flushleft}
\begin{enumerate}[label=\arabic*., ref=\arabic*]
 \setlength{\itemsep}{-1pt}

\item \label{step1} For each column $y \in \{0, \ldots ,
  m-1\}$, and each row $x \in [n]$, the simulator prepares a Bell pair
  $(\ket{00}+\ket{11})/\sqrt{2}$ and outputs half at its outer
  interface.

\item For each column $y \in \{0, \ldots ,
  m-1\}$, and each row $x \in [n]$, the simulator picks an angle
  $\delta_{x,y}\in \{k \pi / 4\}_{k=0}^{7}$ uniformly at random, and
  outputs it at its outer interface. It receives some response
  $s_{x,y} \in \{0,1\}$.

\item The simulator receives $n$ qubits, which correspond to the last
  (output) layer.

\item \label{step2} The simulator transmits all EPR pair half, all angles
  $\delta_{x,y}$, bits $s_{x,y}$ and output qubits to the ideal blind delegated
  quantum computation resource, along with instructions to perform the
  operations described hereafter.
\end{enumerate}

\begin{flushleft}
\textbf{The ideal blind DQC resource}
\vspace{-7pt}
\end{flushleft}
\begin{enumerate}[label=\arabic*., ref=\arabic*]
 \setlength{\itemsep}{-1pt}

\item The blind DQC resource receives the input $\rho_{in}$ and
  a description  of the computation given by angles $\phi_{x,y}$ at its
  $A$\=/interface, and all the information described in
  Step~\ref{step:IR-message} above at its $B$\=/interface.
\item For each $x \in [n]$, it performs the first
  measurement of a teleportation of the input, i.e.,
  for each $x$ it performs a CNOT on the corresponding EPR half using
  the input qubit as control, and measures the EPR half in the
  computational basis. It records the outcome in $i_{x}$.
\item If $i_x = 1$, it updates the measurement angles $\phi_{x,0}$
  and $\phi_{x,1}$.  

\item  For $y = 0, \ldots , m-1$, repeat\\
 \hspace*{\parindent}\hspace*{\parindent}  For $x = 1, \ldots, n$, repeat
\begin{enumerate}
\item It computes the updated measurement
  angle $\phi'_{x,y}$ .
  \item \label{step3} It computes $\theta'_{x,y} = \delta_{x,y} - \phi'_{x,y}$. It
  then applies $Z_{\theta'_{x,y}}$, followed by a Hadamard $H$ and a
  measurement in the computational basis to the $\ith{x}$ qubit of
  the input $\rho_{in}$ if $y = 0$, and to the corresponding EPR half
  if $y > 0$. It stores the result in $r_{x,y}$.
\item If $r_{x,y} = 1$, it flips $s_{x,y}$;
otherwise it does nothing.
\end{enumerate}

\item The ideal blind DQC resource performs the final Pauli
  corrections $ \{Z^{s_{x,m}^{Z} }X^{s_{x,m}^{X}} \}_{x=1}^{n}$ on the
  received output qubits, and outputs the result at its $A$\=/interface.
\end{enumerate}

%\label{prot:UBQC4}
\end{algorithm}

\begin{algorithm}[htbp]

\caption[UBQC with $MRSP_B$ resource and simulator]{$\mathcal{S}_B$ with $MRSP_B$ resource and simulator}
\label{prot:mod}

\begin{flushleft}
\textbf{The simulator}
\vspace{-7pt}
\end{flushleft}
\begin{enumerate}[label=\arabic*., ref=\arabic*]
 \setlength{\itemsep}{-1pt}

\item For each column $y \in \{0, \ldots ,
  m-1\}$, and each row $x \in [n]$, the simulator collects all the measurement operator descriptions, and the states input by Bob. 
\item For each column $y \in \{0, \ldots ,
  m-1\}$, and each row $x \in [n]$, the simulator picks an angle
  $\delta_{x,y}\in \{k \pi / 4\}_{k=0}^{7}$ uniformly at random, and
  outputs it at its outer interface. It receives some response
  $s_{x,y} \in \{0,1\}$.

\item The simulator receives the $n$ bits, which correspond to the last
  (output) layer.

\item \label{step:IR-message} The simulator transmits all the received states, and measurement operator descriptions, all angles
  $\delta_{x,y}$, bits $s_{x,y}$ and output bits to the ideal blind delegated
  quantum computation resource, along with instructions to perform the
  operations described hereafter.
\end{enumerate}

\begin{flushleft}
\textbf{The ideal blind DQC resource}
\vspace{-7pt}
\end{flushleft}
\begin{enumerate}[label=\arabic*., ref=\arabic*]
 \setlength{\itemsep}{-1pt}

\item The blind DQC resource receives the input $\rho_{in}$ and
  a description  of the computation given by angles $\phi_{x,y}$ at its
  $A$\=/interface, and all the information described in
  Step~\ref{step:IR-message} above at its $B$\=/interface.
\item For each $x \in [n]$, it performs the first
  measurement of a teleportation of the input, i.e.,
  for each $x$ it performs a CNOT on the corresponding EPR half using
  the input qubit as control, and measures the EPR half in the
  computational basis. It records the outcome in $i_{x}$.
\item If $i_x = 1$, it updates the measurement angles $\phi_{x,0}$
  and $\phi_{x,1}$.

\item  For $y = 0, \ldots , m-1$, repeat\\
 \hspace*{\parindent}\hspace*{\parindent}  For $x = 1, \ldots, n$, repeat
\begin{enumerate}
\item It computes the updated measurement
  angle $\phi'_{x,y}$ .
\item It computes $\theta'_{x,y} = \delta_{x,y} - \phi'_{x,y}$. It
  then applies the corresponding $\Pi_{\theta'_{x,y}},\Pi_{\theta'_{x,y}+\pi}, $, 
  
  measurement  the $\ith{x}$ system provided by Bob. It stores the result in $r_{x,y}$.
\item If $r_{x,y} = 1$, it flips $s_{x,y}$;
otherwise it does nothing.
\end{enumerate}

\item The ideal blind DQC resource performs the final Pauli
  corrections $ \{X^{s_{x,m}^{X}} \}_{x=1}^{n}$ on the
  received output bits, and outputs the result at its $A$\=/interface.
\end{enumerate}

\label{prot:UBQC4}
\end{algorithm}

We can now ask the question if the UBQC protocol, where Alice has access to $RSP_B$ is also secure, and here is where we immediately profit from the composable framework: since $RSP_B$ can securely implement $MRSP_B,$ UBQC in which Alice has access to $MRSP_B$  also secure, by the composition theorems of AC (or UC). We phrase this as a corollary.

\COR
\label{UBQCprep2}
The UBQC protocol where Alice, instead of access to a quantum channel and a random generator of $\ket{+_{\theta}}$ states, has access to the ideal functionality $RSP_B$ exactly constructs the ideal functionality $\mathcal{S}_B$ - that is, it is perfectly secure. 
\ROC

\end{proof}
%\todo{Proof that correlations are not enough for composability. Not 100\% sure about this... Sketch: 
%If correlations pre-exist, then there is no way for the resource to steer the computation to the correct one. More precisely, if the distinguisher randomly chooses between an honest strategy, and a strategy where he wrecks the computation, the simulator cannot tell the functionality to always output the correct output state. The simulator does not know if the distinguisher played $\pi_B$ honestly - so a situation where the simulator tells the ideal functionality not to output the correct outcome, even though the distinguisher played honestly will occur. In this case, the information the simulator can give to the functionality will not be correlated to the distinguisher's behaviour, since all the simulator outputs is classical information and receives a sequence of random bits. So the resource will not be capable of producing the output which will be realized in the actual protocol. This will easily prove that exact blindness cannot be realized with just initial correlations... For approximate, I will need to think a bit more.}

To illustrate how these definitions and the main theorem can be used, for illustration, we prove the composable security of the so-called two-server protocol introduced in the original UBQC paper \cite{BFK09}.

\subsubsection{Composability of the two-server protocol}
In this subsection we prove the composable security of the two-server protocol defined in the original paper \cite{BFK09}. 
In this protocol, two non-communicating servers share Bell-pairs, and Alice instructs one to measure its halves with respect to random angles. If the first server is honest, this steers the second server's system to a collection of randomly chosen $\ket{+_\theta}$ states. Alice, from the outcomes of the first server's measurements and her instructions knows the states server two is supposed to have. From this point on, Alice continues the basic UBQC protocol with server two.

Formally, we prove that the two-server protocol perfectly securely constructs $RSP_B$ using the resource $\aR$ which is a quantum channel, and the two servers. Formally, we have the following lemma:
\LE
Let $\pi$ be the two-server preparation protocol. Then
\EQ{
\aR,B_1,B_2 \stackrel{\pi, \epsilon=0}{\longrightarrow} RSP_B
}
\EL
\begin{proof}
We model the setting with the promise of no communication between the two servers by assuming that in the real protocol the two non-communicating servers are an untrusted resource with the promises that they cannot communicate, and share an EPR-pair. From the perspective of the distinguisher, this means that the servers have a $D$ interface which the distinguisher can access. $B_1$ has a filtered functionality, a bit $c_1$ which if the distinguisher wishes the server to act malevolently, the distinguisher can pre-set to $1$ (and zero for an honest behavior.) If $c_1=1$ the server awaits the sequence of two-outcome POVM elements, described by pairs of CP maps $\{(\theta, \{\mathcal{E}_\theta^0,\mathcal{E}_\theta^1,\})\}$ which fully describe $B_1$'s behaviour for each $\theta$ he may receive from Alice. $B_2$ also has a filtered functionality, a bit $c_1$ which if the distinguisher wishes the server to act malevolently, the distinguisher can pre-set to $1$. In this case, the server awaits the description of a map $\mathcal{E}_{B_2}$ he will implement on his Bell-pair half before outputting it on his $D$ interface. Server $B_1$ does not output anything. 
The correctness of the protocol, $\emph{i.e.}$  $\pi B_1,B_2\sigma_{\bot}  \aR = RSP_B \sigma_{\bot}$ is  trivial, by the local correctness of the protocol, as proven in \cite{BFK09}.
For the security we need to show the existence of a simulator $\sigma_B$ such that $\pi B_1,B_2  \aR = RSP_B \sigma_{B}$.

Note that the simulator will have a simulated pair of interfaces $B_1$ and $B_2$ toward the distinguisher.
If the distinguisher pre-sets $c_1=0$, the simulator presses $c=0$ on the ideal resource, and collects the output. Then, if the distinguisher pre-set $c_2=0$ on the simulated $B_2$ interface he simply outputs the state he received from the ideal resource. If $c_2=1$ he obtains a description of a map, which he applies on the received state from the ideal resource and outputs that.

The only interesting case is if $c_1=1$, in which case the simulator receives the descriptions of the maps $\{(\theta, \{\mathcal{E}_\theta^0,\mathcal{E}_\theta^1,\})\}$. In this case, the simulator computes the states:
\EQ{
\{\theta,   \sum_{m} \eta^{\theta + m \pi,m}\}}

with 
\EQ{
\eta^{\theta,m} \df Tr_{B1}\left( ({\mathcal{E}_{\theta}^{m}}_{B1} \otimes \Id_{B2}) (\dm{\psi^+}_{B1 B2}) \right)
}
which are the states $B_2$ would have in the real run of the protocol.
These are legal states he can input in the ideal resource, and after he does, he collects the output from the ideal resource. Following this he either just outputs this state to the distinguisher (if $c_2=0$) or additionally applies the map the distinguisher input in the simulated $B_2$ interface, before outputting.
It is easy to see this process is indistinguishable from the real protocol, by construction. 
\end{proof}

In the remainder of the paper, we use the results we have given thus far to show that the capacity for Alice to prepare essentially any two non-orthogonal states, already suffices for blind quantum computation.
We do this by providing protocols which securely construct the functionalities $MRSP_B$ (and $RSP_B$) which use simpler resources.
\section{Security from two non-orthogonal states}
\label{minimal}
In this section, we will provide a protocol in which Alice just needs to be able to prepare two non-orthogonal states, say $\ket{+}$ and $\ket{+_\delta},$ for any angle $0<\delta \not=\pi$, and which constructs the functionality $MRSP_B$. By the results of the previous section, this implies that access to this resource (and, specially to a device which prepares exactly those states) suffices for Alice to run a secure UBQC protocol.

We prove our results through a few intermediary steps. First we will show that
a four state measurement based functionality $MSP_B(4)$ suffices for the exact implementation of $MSP_B(8)$.
Recall, this functionality, when used honestly, prepares the four BB84 states ($\{\ket{+}, \ket{-},\ket{+_{\pi/2}}, \ket{-_{\pi/2}} \}$).
As we will be using variants of the functionalities we defined earlier, but which output all 8 states or just 4, we 
 distinguish between them by putting the state number in parenthesis, so e.g. $MRSP_B(8)$ and $MRSP_B(4)$ differ only in that the first implements the standard 8 states, and the latter the four BB84 states.

Later, we will show that there exist protocol which allow Alice to implement $MRSP_B(4)$, while having just the capacity to implement the two states $\ket{+}, \ket{+_{\pi/2}}.$ 

Finally we generalize this to essentially any two non-orthogonal states.

\subsection{Blindness from 4 states}
First we claim that there exists a protocol $\pi$ such that 
\EQ{
MRSP_B(4) \stackrel{\pi, \epsilon=0}{\longrightarrow} MRSP_B(8).
}
We assume classical channels are always available, as usual.

It will be convenient to assume that when using $MRSP_B(4),$ Alice has the capacity to choose the angles, rather than have them output uniformly at random.

In general, for the functionalities $RSP_B(2N), $ the non-random variants, denoted $SP_B(2N),$ (that is, the variants where Alice chooses her angles), can be efficiently and securely constructed one from the other perfectly.
One direction (`chosen' to `random') has the almost trivial protocol where Alice's part randomly chooses an angle, and Bob's side is trivial. The simulator is trivial as well.
The other direction  $RSP_B(2N) \rightarrow SP_B(2N) $, requires an additional resource: classical channel capable of representing the $2N$ angles. The protocol of Alice accepts an angle as input on the Alice's interface, and the angle from the functionality $RSP_B(2N),$ and sends the difference between the angles to Bob's side. Bob's protocol rotates the received quantum state by the difference. This is easily seen to be a perfectly correct protocol.
The simulator $\sigma_B$ satisfying
\EQ{
\pi_A RSP_B(2N) \mathcal{R}_{c} =SP_B(2N)  \sigma_B,
}
is given as follows.
The $B$ interface has the capacity to collect the description of the measurement operators given by Bob (when $c=1$), along with a quantum system, and to output a quantum system and the description of an angle.  It collects the descriptions $\{\Pi_{\theta}\}_{\theta}$ of the measurement operators, and chooses a random angle $\theta'$. It modifies the descriptions of the operators to  to  $\{\Pi_{\theta-\theta'}\}_{\theta}$ and inputs this into the functionality  $SP_B(2N) .$ If the first set of operators were valid, so is the modified set. It collects the output from $SP_B(2N)$ and sends it to its B-interface (corresponding to the B-interface of the functionality $RSP_B$ used in the real protocol we are simulating), and the angle $\theta'$ , on the part of the interface corresponding to the communication channel $\mathcal{R}_{c}$.
It is clear that the outputs and collective behaviours of both the real and the simulated protocol are the same.
Thus the random, and non-random variants of the preparation functionalities are essentially equivalent. Formally we have
\LE
There exist efficient protocols $\pi$ and $\pi'$ (as given above) such that
\EQ{
SP_B(2N) \stackrel{\pi, \epsilon=0}{\longrightarrow } RSP_B(2N)\\
RSP_B(2N), \mathcal{R}_{c}  \stackrel{\pi', \epsilon=0}{\longrightarrow } SP_B(2N),
}
where $\mathcal{R}_{c} $ is a channel capable of sending one angle.
\EL
For the measurement-based variants of the functionalities, we cannot expect that Alice's interface can accept any angle, as Bob's side can choose the measurements and the measured state. Recall, there the output on Alice's interface depends on a measurement outcome, which may in general be random. 

However, we can guarantee that the outcome be one of two angles: $\theta,\theta+\pi$, thus Alice can choose her angle up to $\pi$.
We will denote such a functionality $MSP_B(2N) $. It receives an angle from Alice   $\theta,$ which directs the functionality to preform the corresponding complete measurement $\Pi_{\theta},\Pi_{\theta+\pi}$ and return the outcome on Alice's interface.

We again have that $MRSP_B(2N)$ can securely construct  $MSP_B(2N) $ (using the same construction for the non-measurement-based variants), but also the reverse, if we use additional classical channels, as follows.

Recall, we are describing a protocol $\pi = (\pi_A, \pi_B)$ such that 
\EQ{
\pi_A MRSP_B(2N) \pi_B = MSP_B(2N)   \bot_B
}
and  the simulator $\sigma_B,$ such that
\EQ{
\pi_A MRSP_B(2N) \mathcal{R}_{c} =MSP_B(2N)  \sigma_B.
}

The protocol proceeds as follows: $\pi_A$ receives a desired angle $\theta$ from its $A-$interface, and collects a random $\theta'$ from the functionality $MRSP_B(4)$. It then uses the channel $ \mathcal{R}_{c}$ to send the angle $\delta = \theta - \theta'$  to the B-side, where $\pi_B$ receives it.
$ \pi_B$ then collects the (honest) state $\ket{+_{\theta'}}$ from the B-interface of $MRSP_B(4),$ applies the $Z_{\delta}$ rotation to it, and outputs the resulting state, which is clearly $\ket{+_\theta},$ which is what $MSP_B$ would output in the same setting.
This proves correctness.

To prove security, we construct the simulator $\sigma_B$ as follows. The $B$ interface has the capacity to collect the description of the measurement operators given by Bob (when $c=1$), along with a quantum system, and to output a quantum system and and angle (from the interface of the channel).  It collects the descriptions $\{\Pi_{\theta}\}_{\theta}$ of the four measurement operators, and chooses a random angle $\theta'$. It modifies the descriptions of the operators to  to  $\{\Pi_{\theta-\theta'}\}_{\theta}$ and inputs this into the functionality  $MSP_B(2N) .$ If the first set of operators were valid, so is the modified set. It collects the output from $MSP_B(2N)$ and sends it to its B-interface (corresponding to the B-interface of the functionality $RSP_B$ used in the real protocol we are simulating), and the angle $\theta'$ (given by two bits), on the part of the interface corresponding to the communication channel $\mathcal{R}_{c}$.
It is clear that the outputs and collective behaviours of both the real and the simulated protocol are the same.
This proves the following lemma:
\LE\label{red1}
There exist efficient protocols $\pi, \pi'$ (as given above) such that
\EQ{
MSP_B(2N) \stackrel{\pi, \epsilon=0}{\longrightarrow } MRSP_B(2N)\\
MRSP_B(2N), \mathcal{R}_{c}  \stackrel{\pi', \epsilon=0}{\longrightarrow } MSP_B(2N),
}
where $\mathcal{R}_{c} $ is a channel capable of sending one angle.
\EL

Then, to show 
\EQ{
MRSP_B(4)  \stackrel{\pi, \epsilon=0}{\longrightarrow} MRSP_B(8),
}
it suffices to show that 
\EQ{
MSP_B(4) \stackrel{\pi, \epsilon=0}{\longrightarrow} MRSP_B(8).
}
where we assume that the classical channels come for free, hence we do not explicitly denote them as available resources.

We now describe the protocol, which requires the functionality $MSP_B(4)$ which prepares the states $\{\ket{+}, \ket{-},\ket{+_{\pi/2}}, \ket{-_{\pi/2}} \} $(when used honestly), and where Alice can choose the bases of the generated states (so chooses between $\ket{+}/\ket{-}$ and $\ket{+_{\pi/2}}/\ket{-_{\pi/2}} $ ) .

We describe and analyze the protocol for the honest case, which will prove correctness, and show security later.
 Details of the protocol are given in Protocol \ref{Protocol:sp4}.

\begin{algorithm}
\caption{Protocol for four states.}
\label{Protocol:sp4}
\label{epsMin}
\begin{itemize}
\item \textit{Input}: No input
\item \textit{Output}: Alice outputs an angle $\theta$ and Bob a qubit in the state $\ket{+_\theta}$, where $\theta$ is chosen uniformly at random in $\{l \pi/4 \vert l = 0 \ldots 7 \}$.

\end{itemize}
\textbf{Protocol steps:}
\begin{itemize}
\item Alice's round:
\begin{enumerate}
\item Alice uses $MSP(4)$, such that it outputs to Bob (for Honest Bob) four qubits $Q_1, Q_2, Q_3, Q_4,$ in the states
\EQ{
Q_1 = Z^{a_1}S^{a_2}\ket{+}\\
Q_2 = Z^{b_1}S^{b_2}\ket{+}\\
Q_3 = Z^{c_1}S^{p}\ket{+}\\
Q_4 = Z^{c_2}S^{1\oplus p}\ket{+}\\
}
where $a_1, b_1, b_2, c_1,c_2 \in_{R}\{0,1\},$ chosen uniformly at random by Alice, and $ a_2, b_2, p$ are chosen uniformly at random by the functionality, and reported to Alice.
\end{enumerate}
\item Bob's round:
\begin{enumerate}
\item Bob applies $S^{\dagger}HS^{\dagger}$ to systems and $Q_3,$ and $Q_4$.
\item Bob applies $Z(\pi/4)$ rotation to system $Q_2$.
\item Bob prepares a fresh system $Q_5$ in the state $\ket{+}.$
\item  Bob performs \ctrlZ entangling and $X$-basis measurement operations: 
\begin{enumerate}
\item Bob entangles $Q_1$ with $Q_3$, measures $Q_1,$ obtaining outcome $o_1$.
 \item Bob entangles $Q_3$ with $Q_5$, measures $Q_3,$ obtaining outcome $o_2$.
 \item Bob entangles $Q_2$ with $Q_4$, measures $Q_2,$ obtaining outcome $o_3$.
  \item Bob entangles $Q_4$ with $Q_5$, measures $Q_4,$ obtaining outcome $o_4$.
\end{enumerate}

\item Bob sends the outcomes $o_1, o_2, o_3$ and $o_4$  to Alice.
\item Bob outputs the system $Q_5$
\end{enumerate}
\item Alice's round:\\
\begin{enumerate}
\item Alice computes 
\EQ{
\theta = (-1)^{c_1 \oplus o_2} (( a_1\oplus o_1 \oplus c_2) \pi + a_2 \pi/2),}
for $p=0$, and
\EQ{
\theta =(-1)^{c_2 \oplus o_4} (( b_1\oplus o_3 \oplus c_1) \pi + b_2 \pi/2 + \pi/4)
}
for $p=1$.
\item Alice outputs $\theta$.

\end{enumerate}
\end{itemize}
\end{algorithm}

Alice first uses 4 calls to $MSP_{B}(4)$ to generate 4 systems, denoted $Q_1, Q_2, Q_3, Q_4,$ which are, in the case of honest Bob, in the states:
\EQ{
Q_1 = Z^{a_1}S^{a_2}\ket{+}\\
Q_2 = Z^{b_1}S^{b_2}\ket{+}\\
Q_3 = Z^{c_1}S^{p}\ket{+}\\
Q_4 = Z^{c_2}S^{1\oplus p}\ket{+}\\
}
where $a_1, a_2, b_1, b_2, c_1,c_2, p \in_{R}\{0,1\},$ chosen uniformly at random by Alice, or by the functionality itself.
Note that the qubits $Q_3$ and $Q_4$ are in correlated states - the operator $S$ chooses the basis, and the two qubits are in two states of opposite bases. But note that the functionality $MSP_{B}(4)$ allows Alice to choose exactly this.
Alice asks Bob to apply $S^{\dagger}HS^{\dagger}$ to systems and $Q_3$ and $Q_4$.
This maps the four possible states, up to global phase, as follows: $(\ket{+}, \ket{-},\ket{+_{\pi/2}}  \ket{-_{\pi/2}}) \stackrel{S^{\dagger}HS^{\dagger}}{\longrightarrow} ( \ket{+}, \ket{-}, \ket{0}, \ket{1}) $.
Alice then asks Bob to apply the $Z(\pi/4)$ rotation to system $Q_2$.
Finally Bob prepares a fresh system $Q_5$ in the state $\ket{+},$ and entangles all the systems using control-Z as follows: $Q_1$ with $Q_3$, $Q_2$ with $Q_4$, and $Q_3$ with $Q_5$ and $Q_4$ with $Q_5$. We thus have two effective computational wires from the input systems $Q_1$ and $Q_2$, leading through connecting qubits $Q_3$ and $Q_4$ to the same system $Q_5.$ However, one of the paths will be broken (depending on the bit $p$) as the Z-eigenbasis states do not get entangled using \ctrlZ\ operations. The same trick is behind the ``trapification'' procedure in the verifiable variant of the UBQC protocol \cite{FK12}.

Bob then measures  all the systems in the Pauli-X basis, obtaining outcomes $o_1, o_2, o_3$ and $o_4$ which are sent to Alice.
If $p=0$, the final qubit is (up to global phase) in the state:
\EQ{
Q_{final} = Z^{c_2} H Z^{c_1 \oplus o_2}H Z^{a_1\oplus o_1}S^{a_2}\ket{+} = X^{c_1 \oplus o_2} Z^{a_1\oplus o_1 \oplus c_2}S^{a_2} \ket{+}.
}

If $p=1$, the final qubit is (up to global phase) in the state:
\EQ{
Q_{final} = Z^{c_1} H Z^{c_2 \oplus o_4}H Z^{b_1\oplus o_3}S^{b_2} Z(\pi/4) \ket{+} = X^{c_2 \oplus o_4} Z^{b_1\oplus o_3 \oplus c_1}S^{b_2} Z(\pi/4) \ket{+},
}
and Alice simply computes this state.
It should be clear that, since all measurement angles are one-time padded by a choice of Alice,  if Alice's bits are chosen uniformly at random, then the resulting states span all $8$ angles. What remains to be seen is that for any choice of responses of Bob, the number of choices of Alice, which are compatible with a fixed final angle $\theta$ is the same for all angles. Only in this case will the distribution over the angles be uniform, which is a requirement of the correctness. To see this holds, first note that the choice of outcomes effectively map the choices of Alice to some other set of choices, bijectively. This implies that if the distribution over final angles is uniform for some sequence of outcomes, than it is uniform for all.
Thus, it suffices to prove the distribution is uniform for the fixed of outcomes $o_j=0,$ for all $j$.
Next, note that it suffices to show that the distribution is uniform for the case $p=0$ (which yield the BB84 states) and, separately that it is uniform for $p=1$ (yielding the same states rotated by $Z_{\pi/4}$).
Note that the expressions for the two cases are essentially the same, and independent except for $c_1$ and $c_2$ which appear in both expressions with exchanged roles.
This implies that it suffices to prove that the distribution is uniform for just one choice, say $p=0$.
We have that $\theta$ is in this case given by
\EQ{
\theta = (-1)^{c_1} (( a_1 \oplus c_2) \pi + a_2 \pi/2) = (-1)^{c_1} (( a_1 \oplus c_2) \pi + a_2 \pi/2).
}
We can immediately see that if we set $c_1=0,$ that the distribution over the angles when we average over $a_1,a_2,c_2$ is uniform, and it is also uniform if  $c_1=1$. Thus, it is uniform if we average over all the parameters.
The same holds in all other cases, so he protocol is perfectly correct.

%Next, we claim that, for any set of measurement outcomes $o_1, o_2, o_3$ and $o_4$, the collection of systems that Alice sends, which yield any final state $\ket{+_\theta}$ or $\ket{-_\theta}$, under honest Bob, is independent from $\theta$.
%Since all the outcomes are one-time padded by (at least) one of the parameters of Alice, the outcomes do not matter, and we can set them to all zero.
%Consider first the four BB84 states ($p=0$).
%where we have two options for $\theta=0$ or $\theta =\pi$. 
%For this case, we find the set of all parameters which yield the $\ket{+}$ ($\theta=0$) or the $\ket{-}$ ($\theta =\pi$) state. Since $p=0,$ and these states are invariant under $X,$ the parameters $c_1,$ $b_1,$ $b_2$ do not influence the final state. This means $Q_2$ (conditioned on the final state being either  $\ket{+}$ or the $\ket{-}$) is maximally mixed. In fact, the only two parameters which are not free are $a_2$ and $p$, which have to be zero.
%However, $a_1$ is free, meaning that $Q_1$ is also maximally mixed. But since $c_1$ and $c_2$ are also free, this means $Q_2$ and $Q_3$ are maximally mixed as well.
%We obtain completely analogous results for all other choices of angles.  This proves correctness.

To prove security, we need to construct the appropriate simulator. The simulator will, for the corrupt case $c=1$, collect 4 sets of descriptions of measurement operators, one for each qubit/system $Q_1, \ldots Q_4$, and collect the systems input by Bob/the distinguisher for all of them (recall the $MSP$ functionalities accept also a system from Bob, and we have one call to them for each system $Q_1$ to $Q_4$).
The overall set of measurement operators input are given with the set
\EQ{
\{ \Pi_{\theta}^{i}\}_{\theta},
}
where for each $i=1,\ldots,4 $ we have that $\Pi^i_{\theta} + \Pi^i_{\theta+\pi} = \mathbbmss{1}$.

Next, consider the function which Alice uses in the real protocol, to compute the realized angle:
If $p=0$ we have that
\EQ{
Q_{final} = X^{c_1 \oplus o_2} Z^{a_1\oplus o_1 \oplus c_2}S^{a_2} \ket{+}} so Alice computes
\EQ{
\theta = (-1)^{c_1 \oplus o_2} (( a_1\oplus o_1 \oplus c_2) \pi + a_2 \pi/2).
}

If $p=1$ we have that
\EQ{
Q_{final} = X^{c_2 \oplus o_4} Z^{b_1\oplus o_3 \oplus c_1}S^{b_2} Z(\pi/4) \ket{+}.
}
so Alice computes
\EQ{
\theta =(-1)^{c_2 \oplus o_4} (( b_1\oplus o_3 \oplus c_1) \pi + b_2 \pi/2 + \pi/4).
}

We can abstract Alice's computation as the function $f(a_1,a_2,b_1,b_2,c_1,c_2,p,o_1,o_2,o_3,o_4) \in \{ l \pi/4\}_{l=0}^7,$ which generates the Alice's computed angle, and given by the formulae above.

The simulator proceeds as follows: it collects all the descriptions, the systems Bob inputs, and the measurement outcomes $(o_1,o_2,o_3,o_4)$ which are all input on Bob's interface (as in the real protocol).
Then, for each $\theta$ in the set of all angles, it computes the set $s(\theta) = \{(a_1,a_2,b_1,b_2,c_1,c_2,p) | f(a_1,a_2,b_1,b_2,c_1,c_2,p,o_1,o_2,o_3,o_4) = \theta\},$
where this set depends on Bob's reported outcomes.
Note that the bits $a_1,a_2$ parametrize the measurement operators for first system, $b_1,b_2$ for the second, and the three bits $c_1,c_2,p,$ characterize the last two systems as follows:
for the first two systems we have that $\Pi^1(a_1,a_2) = \Pi^{1}_{(a_1\pi + a_2\pi/2)}$, and $\Pi^2(b_1,b_2) = \Pi^{2}_{(b_1\pi + b_2\pi/2)}$.
For the third and fourth we have that:
\EQ{
\Pi^{3,4}(c_1,c_2,p) = \Pi^{3}_{c_1\pi+ p\pi/2}\otimes\Pi^4_{c_2\pi+(1\oplus p) \pi/2}.
}
We now claim that the operators
\EQ{
\Pi_{\theta} = \sum_{(a_1,a_2,b_1,b_2,c_1,c_2,p) \in s(\theta)}\Pi^1(a_1,a_2) \otimes \Pi^2(b_1,b_2) \otimes \Pi^{3,4}(c_1,c_2,p)
}
form a valid input for the functionality $MRSP_B(8),$ in other words, that for all $(o_1,o_2,o_3,o_4)$ 
\EQ{
\Pi_{\theta} + \Pi_{\theta+\pi} = \mathbbmss{1}. \label{crit11}
}
We, however, already know that for all fixed $a_2,b_2, c_1, c_2$, and for all outcomes  $(o_1,o_2,o_3,o_4)$ it holds that 
\EQ{
\sum_{a_1} \Pi^1(a_1,a_2)  = \mathbbmss{1}; \sum_{b_1} \Pi^2(b_1,b_2)  = \mathbbmss{1},
}
as they are valid inputs for the functionalities $MSP_B(4). $ But it also holds that for any fixed $p$
\EQ{
\sum_{c_1,c_2 }\Pi^{3,4}(c_1,c_2,p) =  \left(\sum_{c_1}\Pi^{3}_{c_1\pi+ p\pi/2} \right)\otimes  \left(\sum_{c_2}\Pi^4_{c_2\pi+(1\oplus p) \pi/2} \right) = \mathbbmss{1} \otimes  \mathbbmss{1},
}
for the same reason.

Finally, by the analysis of the frequencies of $(a_1,a_2,b_1,b_2,c_1,c_2)$, compatible with an angle in the even parity angles $\{ 2l \pi/4 \}_{l=0}^3$ (case $p=0$) and odd parity case (case $p=0$) $\{ (2l+1) \pi/4 \}_{l=0}^3$,
we have that the cases where $a_1, b_1, c_1, c_1$ equal 1, and cases where they equal zero, appear the same number of times, if we constrain $f(a_1,a_2,b_1,b_2,c_1,c_2,p,o_1,o_2,o_3,o_4) = \theta$ or  $f(a_1,a_2,b_1,b_2,c_1,c_2,p,o_1,o_2,o_3,o_4) = \theta+\pi,$ for any $\theta$, and any fixed choice of outcomes $o_j.$ 
To see this, note that fixing  $\theta$ only fixes $a_2, b_2$ and $p$ and other parameters, in particular $a_1,a_2,c_1 $ and $c_2$ are freely averaged over. 

Thus the claim in Eq. (\ref{crit11}) holds. 
But then the simulator can input exactly the measurement operators 
\EQ{
\Pi_{\theta} = \sum_{(a_1,a_2,b_1,b_2,c_1,c_2,p) \in s(\theta)}\Pi^1(a_1,a_2) \otimes \Pi^2(b_1,b_2) \otimes \Pi^{3,4}(c_1,c_2,p)
}
into the functionality $MRSP_B(8),$ along with all the system it collected from its Bob-interface.
The outcomes will, by construction, result in the same state as in the real protocol, which proves the security with zero error.
Thus, we have proven the following lemma.
\LE
There exist an efficient protocol $\pi$ (as given above) such that
\EQ{
MSP_B(4) \stackrel{\pi, \epsilon=0}{\longrightarrow } MRSP_B(8).
}

\EL

By the previous reduction of Lemma \ref{red1}, this also means that
$MRSP_B(4)$ can be used to exactly construct $MRSP_B(8),$ which will be important in the next section.

We point out that since $SP_B(4) \rightarrow RSP_B(4) \rightarrow MRSP_B(4),$ this also means that having the capacity to prepare just the four states suffices for blind quantum computing with zero error. The protocol is exactly the same as the protocol we provided above.
There have already been protocol proposed which achieve blind quantum computing using just four states 
\cite{GMMR13,FBSYLPJR14}. Next, we show how to realize $MRSP_B(4)$ securely using just two states, which we believe is a new result.

\subsection{Blindness from 2 states}

Assume that Alice has a device, resource $SP(2)$ which randomly chooses between two states $\ket{+}$ and $\ket{+_{\pi/2}}$ and sends them to Bob.
We now give a protocol which, using  $SP(2)$ many ($N=2K$) times, constructs  $RSP_{B}(4)$ securely, but for the first time with finite error $\epsilon$.

This protocol is essentially a one-dimensional measurement based computation, where we make sure that the qubits never leave the XY plane of the Bloch sphere. The qubits are sequentially enangled, and one of the two is measured.
This can be done such that the resulting qubit is characterized by an angle which is the sum (up to signs which depend on the measurement outcomes) of the angles of the two qubits. When this is done sequentially, the final qubit is in a state whose characterizing angle is (up to signs) the sum of all the angles which appeared in the initial sequence. The initial angles are chosen uniformly at random by Alice, and this process constitutes a random walk on the unit circle of angles. This walk quickly approaches a uniform distribution over all the angles which can be reached.
The details of the protocol are given in Protocol \ref{prot21}. 

\begin{algorithm}
\caption{Protocol for two states}
\label{prot21}
\label{epsMin}
\begin{itemize}
\item \textit{Input}: Security parameter $N=2K$.
\item \textit{Output}: Alice outputs an angle $\theta$ and Bob a qubit in the state $\ket{+_\theta}$, where $\theta$ is chosen uniformly at random in $\{l \pi/2 \vert l = 0 \ldots 3 \}$.

\end{itemize}
\textbf{Protocol steps:}
\begin{itemize}
\item Alice's round:
\begin{enumerate}
\item Alice uses $SP(2)$ (outputting $\ket{\psi(i)_{j}} = 1/2(\ket{0}+ e^{i\pi/2}  \ket{1})$, to Bob and the bit $i$ to Alice) and generates $N$ states, stores their encoding $\{ i_j\}_{j=1}^N$.
\end{enumerate}
\item Bob's round:
\begin{enumerate}
\item Bob applies the $Z_{-\pi/4}$ rotation to each qubit.
\item Bob applies the 1DQC subroutine \cite{DKL12} to the qubits $\ket{\psi(i_j)_j},$ for $j=1,\ldots,N$ given by the following:
\item 
\label{bobepsMin} For each $j=1,\ldots,N-1$\\
\begin{itemize}
\item Bob applies 
a Hadamard to the $j^{th}$ qubit.
\item Bob $ctrl-$Z to qubit $j$ and $j+1$.
\item He measures the $j^{th}$ qubit with respect to the $X$ basis. \item He stores the output in the bit $b_j$, discards the measured qubit, and applies $X^{b_j}$ to the $(j+1)^{st}$ qubit.
\end{itemize}
\item \label{corrections} From the bitstring $b_1, \ldots, b_{N-1}$, Bob computes the bitsting $t_k = \bigoplus_{j<k} b_k$  ($\oplus$ denotes mod 2 addition) and sends this bitstring to Alice.
\end{enumerate}
\item Alice's round:\\
\begin{enumerate}
\item Alice computes 
\EQ{
\theta = \sum_{j=1}^{N-1} (-1)^{t_j \oplus i_j} \pi/4 \ \mod 2\pi
}
\item Alice outputs $\theta$
\end{enumerate}
\end{itemize}
\end{algorithm}

For this protocol we claim that
\EQ{
\pi_A SP(2) \pi_B \approx_{\epsilon_{corr}} MRSP_{B}(4)   \bot_B
}
and that there exists a simulator $\sigma_B,$ such that
\EQ{
\pi_A SP(2) \mathcal{R}_{2c} \approx_{\epsilon_{sec}} MRSP_B(4)  \sigma_B.
}

We begin with the proof of correctness, and in the process establish some properties which will help us construct the simulator.
To prove the $\epsilon-$correctness of Protocol \ref{prot21}, since there are no inputs to the protocol, it will suffice to show that the state $\sigma_{AB}^{real}$ generated by utilizing the resource $\pi_A SP(2) \pi_B$  is $\epsilon$-close to the state $\sigma_{AB}^{ideal}$ generated by one use of the resource $MRSP_{B}(4)   \bot_B,$ which is given with
\EQ{
\sigma_{AB}^{ideal} = \sum_{\theta} \dfrac{1}{4} \dm{\theta}_A \otimes \dm{+_{\theta}}
}

We construct the state $ \sigma_{AB}^{real}$, by going through the steps of the protocol. 
The protocol consists of 4 interactions between Alice and Bob. The state of Alice and Bob after Alice's round, and Bob's local rotation, is given with:
\EQ{
\sigma_{AB}^{1} = 2^{-N}\sum_{\ora{i}} \dm{\ora{i}}_A \otimes \dm{\psi{(\ora{i})}},
}
where with $\ora{i}$ we denote the entire sequence of bits $N$ specifying the state $\ket{\psi(\ora{i})}\mathop{:}= \otimes_{j=1}^{N} \ket{\psi(i_j)}$.

In the following round, Bob applies the 1DQC sub-protocol \cite{DKL12}, which outputs $N-1$ measurement outcomes $b_i$. From the outcomes, he computes the local corrections $t_i$ (Step \ref{corrections} in the protocol), and sends them to Alice. He traces out the correction bits.   
The joint state of Alice and Bob is, after these steps given with

\EQ{
\sigma_{AB}^{2} =2^{-2N} \sum_{\ora{i},\ora{t}} \dm{\ora{i}} \otimes \dm{\ora{t}}_A  \otimes \dm{+_{\theta(\ora{i}, \ora{t})}} }.

The angle $\theta(\ora{i},\ora{t})$ is given by the expression
\EQ{
\theta(\ora{i}, \ora{t}) =  \sum_{j=1}^{N} (-1)^{ i_j \oplus t_j} \dfrac{\pi}{4} \mod 2\pi,
}
where we take that $t_N=0$.
Then, Alice computes the angle $\theta$ using the expression above, which we denote as the function $f(\ora{i}, \ora{t}) = \theta(\ora{i}, \ora{t}), $ to clearly distinguish it from the angle itself.

Here, note that we have demanded that the total number of systems sent be even ($N=2K$). This implies that we have a sum of an even number of angles $\pi/4,$ some of which come with a negative sign. Since the total is positive, then either both the number of the plus signs and the minus signs is even, or both are odd.
Either way, the difference of their numbers is even. This implies the angle $\theta$ will always be one of the four angles $\{k \pi/2 \}_{k=0}^3$. Note that if we had not introduced the $Z_{-\pi/4}$ rotation in the protocol, the end state would always be in the set of four angles, but the influence of the corrections on the final angle would be slightly more difficult to analyse - but since this is just a local unitary, if the protocol with the $Z_{\pi/4}$ rotation is secure, so is the one without it.

Following this, Alice computes her angle value $\theta$ using the function $f$, traces out the irrelevant information, and obtains  

\EQ{
\sigma_{AB}^{real} = \sum_{\theta} p(\theta)\ \dm{\theta}_A \otimes  \dm{+_{\theta}} .
}

The probabilities $ p(\theta)$ are not perfectly uniform - they correspond to the fractions of sequences $(i_1\oplus t_1, \ldots, i_N \oplus t_n)$ which map to a given angle $\theta$ under $f$. 
If we fix an arbitrary sequence of corrections $\ora{t}$, and introduce the  $i'_j = i_j \oplus t_j$, we have that 
\EQ{
f(\ora{i}, \ora{t} ) = f(\ora{i'},0) = \sum_{j=1}^{N} (-1)^{i'_j} \dfrac{\pi}{4} \mod 2\pi.
}
For simplicity, we write $f(\ora{i'}) \mathop{:}=  f(\ora{i'},0)$
Note that since all sequences $\ora{i'}$ appear in the joint state, for a fixed set of corrections, we have that $p(\theta) = |f^{-1}(\ora{i'}) |/2^{N},$ where $|f^{-1}(\ora{i'})| $ is the size of the set $f^{-1}(\ora{i'}) = \{ \ora{i'} | f^{-1}(\ora{i'}) ) = \theta\}.$

To specify this distribution we now explicitly compute the sizes of these sets for all four angles, and to make this as simple as possible, we reshuffle the expressions a bit.
We have that 
\EQ{
 f(\ora{i'})  = \sum_{j=1}^{N} (-1)^{i'_j} \dfrac{\pi}{4} \mod 2\pi\\
 f(\ora{i'}) + N\pi/4 \mod 2\pi  = \sum_{j=1}^{N} (-1)^{i'_j} \dfrac{\pi}{4}  + N\pi/4  \mod 2\pi\\
  f(\ora{i'}) + N\pi/4 \mod 2\pi  = \sum_{j=1}^{N} ( (-1)^{i'_j} \dfrac{\pi}{4} + \pi/4)  \mod 2\pi\\
 f(\ora{i'}) + N\pi/4 \mod 2\pi  = \sum_{j=1}^{N} ( (1-i'_j) \pi/2)  \mod 2\pi\\
  f(\ora{i'}) + N\pi/4 \mod 2\pi  =N \pi/2 - \sum_{j=1}^{N}i'_j \pi/2  \mod 2\pi\\
    f(\ora{i'}) - N\pi/4 \mod 2\pi = -  \sum_{j=1}^{N}i'_j \pi/2  \mod 2\pi\\
    K\pi/2-  f(\ora{i'})    \mod 2\pi =   \sum_{j=1}^{N}i'_j \pi/2  \mod 2\pi\\
}
Note that, in the left-hand side of the expression in the last line, we simply have a permutation of the angle given with $\theta \rightarrow K\pi/2 - \theta,$ which depends on the choice of $K$. The probability of Alice computing the angle $l \pi/2 = \theta  =  K\pi/2-  \theta'$ is then given by the number of sequences $\ora{i'}$ such that $\sum_j i'_j \mod 4 = l$, divided by the total number. If we, for the moment, assume that $K$ is even as well, so $N=4M$, this is given by the following expressions:
\EQ{
p(0) = \dfrac{\sum_{h=0}^{M} \binom{4 M}{4h }}{2^{4M}}, \textup{for}\ l=0, \\ 
p(l \pi/2) = \dfrac{\sum_{h=0}^{M-1} \binom{4 M}{4h+l }}{2^{4M}} \textup{for}\ l>0.
}
These can be evaluated to closed form expressions:
\EQ{
p(\pi/2) = p(3\pi/2) = 1/4\\
p(0) = 1/4 + (-1)^{M}\ 2^{- K-1}\\
p(\pi) =  1/4 + (-1)^{M+1}\ 2^{- K-1}.
}
The odd-parity angles ($\pi/2 , 3\pi/2$) have exactly 1/4 probability of appearing, and the other two converge to 1/4 exponentially quickly in $K=N/2.$

Form the above, we have that  $\epsilon_{corr} \leq 1/2 \sum_{\theta} | p(\theta) - 1/4| = 2^{- 2 K-1} = 2^{- N/2-1} $. In the analysis we have assumed that the number of the systems Alice uses  is of the form $N=4M,$ but in the case it is not, the decay is still exponential for any $N$, as the distance between the distributions is upper bounded by $2^{- N'/2-1},$ for the largest $N' \leq N,$ which is of the form $N' = 4M$. This establishes the correctness within $\epsilon_{corr} \leq 2^{- N/2-1}.$

For the security of the protocol, we need to construct a corresponding simulator for which the expression
\EQ{
\pi_A SP_B(2) \mathcal{R}_{2c} \approx_{\epsilon_{sec}} MRSP_B(4)  \sigma_B
}
holds.

To evaluate the distinguishing advantage $\epsilon_{sec}$ above, we will consider the structure of the states Bob has in the real protocol ( after using $\pi_A SP_B(2) \mathcal{R}_{2c} $), and the simulated protocol (after using $MRSP_B(4)  \sigma_B$). 
Since the protocol has only two rounds and takes no input on Alice's interface, the set of distinguisher's strategies consist of only of the choice of maps he applies (pretending to be Bob, at Bob's round), followed by trying to identify which (real or simulated) state he obtains. 
Before we can do this, we must first provide the simulator, and the basic idea is to again use a steering-type approach, but this time approximate. 
First, we will consider again joint state after a run of the real protocol, and re-write in a form which will make the suitable simulator apparent. Following this, we will prove a couple of relevant properties of the states Alice sends in the real protocol, which we will finally use to bound the distance between the states realized by the real and simulated protocol.

Consider now the state joint state output by the real protocol, where we assume the distinguisher (on Bob's side) applies any collection of CP (and trace non-increasing) maps $\{\mathcal{E}^\ora{t} \}_{\ora{t}}$ which are a resolution of a CPTP map. Note, this is all the distinguisher can do at this point.
We obtain
\EQ{
\sigma_{AB}^{real} =\sum_{\theta} \dm{\theta}_A \otimes   
 \sum_{\ora{t}} \mathcal{E}^{\ora{t}}    \left( \dfrac{1}{|\{\ora{i} | f(\ora{i}, \ora{t}) = \theta \}|}  \sum_{\ora{i}, s.t. f(\ora{i},\ora{t)} = \theta}   \dm{\psi(\ora{i})}      \right) .
}

Next, note that 
\EQ{
 | \{ \ora{i}| f(\ora{i}, \ora{t}) = \theta \}| =  | \{ \ora{i}| f(\ora{i}\oplus \ora{t},0 ) = \theta \}|  =  | \{ \ora{i}  | f(\ora{i}, 0) = \theta \}|  =\mathop{:} f^{-1}(\theta),
 }
as $\ora{i} \rightarrow \ora{i} \oplus \ora{t}$ is a bijection between the sets in the second equality, and the first equality holds by the definition of $f$.
Next, let $V(r),$ be the identity for $r=0$ and the unitary map flipping between $\ket{+}$ and $\ket{+_{\pi/2}}$ (up to global phase), which is essentially the reflector about $\ket{+_{\pi/4}}$.
Let $U(\ora{t}) := \otimes_{j=1}^{N} V(t_j).$
Now we have that 
\EQ{
 \sum_{\ora{i}, s.t. f(\ora{i},\ora{t}) = \theta}   \dm{\psi(\ora{i})}    =  \sum_{\ora{i}, s.t. f(\ora{i},0) = \theta}   \dm{\psi(\ora{i} \oplus \ora{t})} =   \sum_{\ora{i}, s.t. f(\ora{i},0) = \theta} U(\ora{t})  \dm{\psi(\ora{i} )} U(\ora{t})^{\dagger}.
}
Thus we can re-write the final state as
\EQ{
\sigma_{AB}^{real} =\sum_{\theta} \dm{\theta}_A \otimes   
 \sum_{\ora{t}} \mathcal{F}^{\ora{t}}    \left( \dfrac{1}{|f^{-1}(\theta)|}  \sum_{\ora{i}, s.t. f(\ora{i}) = \theta}   \dm{\psi(\ora{i})}      \right) ,
}
where $\mathcal{F}^{\ora{t}} = \mathcal{E}^{\ora{t}} \circ U(\ora{t}), $ where the latter acts as $ \mathcal{E}^{\ora{t}} \circ U(\ora{t}) (\rho) = \mathcal{E}^{\ora{t}}(U(\ora{t}) \rho U(\ora{t})^\dagger)  $ .

Let $\xi(\theta) =  \dfrac{1}{|f^{-1}(\theta)|}  \sum_{\ora{i}, s.t. f(\ora{i}) = \theta}   \dm{\psi(\ora{i})}   $ (which is a normalized state),
and let \EQ{\chi(\theta) = 1/2(\xi(\theta) + \xi(\theta+\pi)).}
Now, we can apply Lemma \ref{steer} to represent the states $\xi(\theta)$ as the states of one partition of a bipartite state, conditional on a measurement outcome of a measurement of the other partition. Note that we only have two states $\chi(\theta)$ (as we have four angles, and these states are averages over states characterized by the opposite angles).
In other words, the state $\chi(\theta)$ is characterized solely by the parity of the angle $\theta,$ defined as follows: if we represent the angle using two bits:  $\theta=b_1 \pi + b_2 \pi/2,$ then the bit $b_2$ is the parity of the angle $\theta$.
Thus we can also denote these states
\EQ{
\chi(0) \mathop{:}= \chi(0\times \pi) = \chi(\pi);\ \chi(1)\mathop{:} = \chi(\pi/2) =   \chi(3\pi/2).
}
Let $\ket{\chi(p)}_{12}$ be the  purification of the state $\chi(p),$  (for $p=0,1$) as in Lemma \ref{steer}.
Then Lemma \ref{steer} provides a pair of pairs of complete measurement operators $\Pi(\theta), \Pi(\theta + \pi)$ (the descriptions of which require only the descriptions of the states $\chi(p)$ and $\xi(\theta)$),
such that 
\EQ{
\xi(\theta) =2Tr_{1}\left[ (\Pi(\theta)_{1} \otimes\mathbbmss{1}_{2}) \dm{ \chi(\theta)}_{12} \right].
}
Note that we also only have two pairs of measurement operators, so if we denote the parity of the angle with $\oplus(\theta)$ the joint state can be written as:

 \EQ{
\sigma_{AB}^{real} =\sum_{p=0}^{1} \sum_{{\theta\ s.t.,\atop \oplus{\theta} = p}} \dm{\theta}_A \otimes   
 \sum_{\ora{t}} \mathcal{F}^{\ora{t}} \left( 2Tr_{1}\left[ (\Pi(\theta)_{1} \otimes\mathbbmss{1}_{2}) \dm{ \chi(p)}_{12} \right] \right),
}
 
where we have broken the sum over the angles into the sums over even and odd angle parities.
The expression above is indicative to what the simulator should do. The simulator will prepare an entangled state, and output one part on his Bob interface, and the other part it will keep. It will collect the outcomes $\ora{t},$ and submit exactly the measurement operators $\{\Pi(\theta)\}_\theta$ to the ideal functionality.
As we will show presently, the states  $\dm{ \chi(p)}_{12} $ are exponentially close to some two states
$\ket{\eta'(p)}_{12},$ which both purify the mixture over all states Alice may send. 
There are two technicalities we will have to sort out. First, the simulator does not know $p$ as it is chosen by the ideal resource - it can only prepare one state. Second, in the derivations of the real state, we have also subsumed the maps $U(\ora{t})$ which we still have to take into account in the simulated protocol. 

Both problems can be resolved by noting that two purifications of the same local state (the one output on Bob's interface, in our case), are always equivalent up to local unitaries (if the purifying systems have the same dimensions).
This means that the simulator can prepare any state $\ket{\eta}_{12}$ which purifies the state Alice sends in the real protocol, and deal with the correcting local unitaries later.

We define then
\EQ{
\ket{\eta}_{12} = 2^{-N}\sum_{\ora{i}} \ket{i} \otimes \ket{\psi(i)},
}
and note that the reduced state $\eta_2$ is just a uniform mixture over all the states Alice may send.

Now, as we show momentarily, there exist two other purifications of  $\eta_2$, denoted $\ket{\eta(p)}_{12}$ which are close to the states $\dm{ \chi(p)}_{12},$ (for matching $p$). But then there exists two local unitaries, $W(0), W(1)$ on the sub-system $1$ such that
\EQ{
\ket{\eta(p)}_{12} = (W(p)_1\otimes \mathbbmss{1}_2) \ket{\eta}_{12}.
}

The simulator does not have the information to know which $W(p)$ to apply, but this is not a problem, as these can be subsumed in the descriptions of the measurement operators the simulator inputs to the ideal resource.

The last thing we need to take care of are the local unitaries $U(\mathbf{t}),$ which act on system $2$ (in the  derivation of the state of the real protocol), and which depend on the outcomes of Bob. By the same arguments, this too can be circumvented by applying a local operation on system $1,$ after Bob/distinguisher sends the outcomes $\mathbf{t}$, as we have:

\EQ{
2^{-N} (\mathbbmss{1}\otimes U(\ora{t}))\sum_{\ora{i}} \ket{i} \otimes \ket{\psi(i)} = 2^{-N}({X(\ora{t})}\otimes \mathbbmss{1})\sum_{\ora{i}} \ket{i} \otimes \ket{\psi(i)},
}
where $X(\ora{t}) = \otimes_{j=1}^N X^{t_j},$ and $X$ is the Pauli sigma-$X$ operator. 

Combining the two resolutions (for $W(p)$ and $U(\mathbf{t})$), we have the following behaviour of the simulator.
The simulator prepares the state $\ket{\eta}_{12},$ and outputs subsystem $2$. It collects the responses $\mathbf{t}$ from its Bob-interface, and submits the system in partition $1$ to the simulator, along with the following description of measurement operators:
\EQ{
\Pi(\theta, \ora{t}) = W(\oplus\theta)X(\ora{t})  \Pi(\theta)X(\ora{t}) W(\oplus\theta)^{\dagger}.
}
We have still not specified the unitaries $W(p),$ and we shall do this last. Before doing so, we will now simplify the expression for the trace distance of the final states of the real and simulated protocol, where the simulator is given as above.
By the construction of the measurement operators the simulator provides, we immediately have that the state of the simulated protocol is given with:

 \EQ{
\sigma_{AB}^{simulated} =\sum_{p=0}^{1} \sum_{\theta, \oplus{\theta} = p} \dm{\theta}_A \otimes   
 \sum_{\ora{t}} \mathcal{F}^{\ora{t}} \left( 2Tr_{1}\left[ (\Pi(\theta)_{1} \otimes\mathbbmss{1}_{2}) \dm{ \eta(p)}_{12} \right] \right) .
}

We now compute $\Delta = 1/2 || \sigma_{AB}^{simulated} - \sigma_{AB}^{real}||.$ Since the states in Alice's register are orthogonal, the distance breaks immediately into the sum over distances, and what remains is:
\EQ{
2\Delta \leq \sum_{p=0}^{1} \sum_{\theta, \oplus{\theta} = p} \left\|   \sum_{\ora{t}} \mathcal{F}^{\ora{t}} \left( 2Tr_{1}\left[ (\Pi(\theta)_{1} \otimes\mathbbmss{1}_{2}) \left( \dm{ \eta(p)}_{12}  - \dm{ \chi(p)}_{12} \right)   \right] \right)   \right\|.
}
 
 Next, we can note that the partial trace commutes with the CP maps, and also that the measurement $\Pi(\theta)$ can be absorbed into the maps (making them also angle dependent, which we denote with $\mathcal{F(\theta)}^{\ora{t}} $), so we obtain:
 \EQ{
\Delta \leq \sum_{p=0}^{1} \sum_{\theta, \oplus{\theta} = p} \left\|   \sum_{\ora{t}} \mathcal{F(\theta)}^{\ora{t}} \left( \dm{ \eta(p)}_{12}  - \dm{ \chi(p)}_{12} \right)   \right\| =\\
\sum_{\theta, \oplus{\theta} = 0} \left\|   \sum_{\ora{t}} \mathcal{F(\theta)}^{\ora{t}} \left( \dm{ \eta(0)}_{12}  - \dm{ \chi(0)}_{12} \right)   \right\| +\\+ \sum_{\theta, \oplus{\theta} = 1} \left\|   \sum_{\ora{t}} \mathcal{F(\theta)}^{\ora{t}} \left( \dm{ \eta(1)}_{12}  - \dm{ \chi(1)}_{12} \right)   \right\|
}
 Next, note that we can sum the maps over $\ora{t},$ still obtaining CP maps. Moreover, these maps are trace non-increasing, thus they are contractive relative to the trace norm. Overall, we get

  \EQ{
\Delta 
 \leq2 \left\|  \dm{ \eta(0)}_{12}  - \dm{ \chi(0)}_{12}   \right\| + 2 \left\|    \dm{ \eta(1)}_{12}  - \dm{ \chi(1)}_{12}   \right\|.
\label{overall}}
What remains is to define the states $\ket{\eta(p)}$ (which implicitly specifies the map $W(p)$ as well) and bound the distances.

To do so, it will suffice to prove that the mixture over all states Alice may send is $\epsilon'$ close to the state $Tr_2[\dm{ \chi(p)}_{12} ],$ for both $p$ .
Then, by Uhlmann's theorem, there exist purifications ($\ket{\eta(p)}_{12}$) which are no more than $2\sqrt{\epsilon'}$ far from the states $\dm{ \chi(p)}_{12}$, for the respective values of $p$.

First, define the states which are the even and odd parity mixtures:
\EQ{
\psi_p = 2^{1-N} \sum_{\ora{i} | |\ora{i}|\!\!\! \mod 2 = p} \dm{\psi(\ora{i})}.
}

Next, by the results of the paper \cite{Fuchs99}, Section 7, if the states Alice sent were either $\cos(\alpha) \ket{0} + \sin(\alpha) \ket{1}$ or $\cos(\alpha) \ket{0} -\sin(\alpha) \ket{1}$, then the trace distance $\delta$ between the state which is a mixture of all odd parity states, and the state which is a mixture of all even parity states is bounded by
\EQ{ \delta \leq|\sin(2 \alpha)|^N \label{distance}.}
These states ($\cos(\alpha) \ket{0} \pm \sin(\alpha) \ket{1}$ ) are unitarily equivalent to the states Alice sends in our protocol for the choice $\alpha=\pi/8.$ This gives the trace distance $\epsilon' \leq 2^{-N/2}$ between $\psi_{0}$ (mixture of all even parity states Alice sends) and $\psi_{1}$ (mixture of all even parity states).

Next, note that for the mixture over all states $\eta = 1/2(\psi_{0} + \psi_{1})$ we have that
\EQ{
1/2|| \eta - \psi_{0}|| =  1/2|| \eta - \psi_{1}||  =1/4 ||    \psi_{1} - \psi_{0}|| \leq \dfrac{1}{2} \epsilon'.
}
Next, note that the state $\chi(1) = Tr_1[ \dm{\chi(1)}_{12}] $ is exactly the state $\psi_1,$ so we immediately have one part of the claim: the state $\eta$ is $ 1/2 \epsilon'$ close to $\chi{1}$.

For the state $\chi(0) = Tr_1[ \dm{\chi(0)}_{12}] $ we almost have the same claim, but not quite.

Recall that  $\chi(0)$ is given with
\EQ{
\chi(0) = \dfrac{1}{2|f^{-1}(0)|}  \sum_{\ora{i}, s.t. f(\ora{i}) = 0}   \dm{\psi(\ora{i})}  +   \dfrac{1}{2|f^{-1}(\pi/2)|}  \sum_{\ora{i}, s.t. f(\ora{i}) = \pi}   \dm{\psi(\ora{i})} 
}
which is not exactly the uniform mixture, as one of the weights is slightly larger then the other. This was also the reason the probabilities in the correctness part of the proof were not exactly uniform.
To give the distance between $\chi(0)$ and $\eta$, we evaluate the distance between  $\chi(0)$ and $\psi_0$, and we will have our claim by the triangle inequality.

We have that
\EQ{
 1/2 || \chi(0) - \psi_0 || = 1/2\left\| \left( \dfrac{1}{2| f^{-1}(0) |} - 2^{-N+1} \right) 
   \sum_{\ora{i}, s.t. f(\ora{i}) = 0}   \dm{\psi(\ora{i})}  +  \right. \\ \left. 
 \left( \dfrac{1}{2| f^{-1}(\pi) | }-  2^{-N+1}\right) 
   \sum_{\ora{i}, s.t. f(\ora{i}) = \pi}   \dm{\psi(\ora{i})} 
 \right\| 
 }
Recall that, when we assume $N=4M$,  for $N(l) = | \{\ora{i}|f(\ora{i}) = l\pi/2 \}| $ we have that 
\EQ{
A(0) = \sum_{h=0}^{M} \binom{4 M}{4h },\ \textup{for}\ l=0, \\ 
A(l ) = \sum_{h=0}^{M-1} \binom{4 M}{4h+l },\ \textup{for}\ l>0,
}
which we computed (implicitly) in the calculations of the probabilities $p(\theta).$
Thus for the cases $\theta = r\pi,$ (for $r\in \{ 0,1\}$) we get
\EQ{
 1/2 || \psi_0 -\chi(0) ||  \leq \sum_{r=0}^{1}   \dfrac{2^{-N+2}}{2^{3M}+(-1)^{M+r} } \left\|   \sum_{\ora{i}, s.t. f(\ora{i}) = r\pi}   \dm{\psi(\ora{i})} \right\| = \\
  \sum_{r=0}^{1}   \dfrac{2^{-N+2}}{2^{3M}+(-1)^{M+r} } (2^{N - 2} + (-1)^{M+r}\ 2^{2 M-1}) \leq 2^{-3N/4+2}.
}
Again, we have assumed that $N$ is of the form $N=4M,$ but the scaling remains the same even if we do not.

Thus we have that $ 1/2 ||\psi_0- \chi(0) || \leq 2^{-3N/4+2}. $ 
Since we have already seen that  $1/2|| \eta - \psi_0|| \leq 1/2 \epsilon'$,  we have the following bound

\EQ{
1/2 || \eta - \chi(0) || \leq 2^{-3N/4+2}+  \epsilon'/2 =\mathop{:} \epsilon''.
}
Then, by Uhlmann's theorem, and applying it to both parity cases, there exists purifications $\{ \ket{\eta(p)}\}_p$ such that 
\EQ{
1/2|| \dm{\eta(0)}_{1,2} - \dm{ \chi(0)}_{1,2}||\leq 2\sqrt{\epsilon''}, \\
1/2|| \dm{\eta(1)}_{1,2} - \dm{ \chi(1)}_{1,2}||\leq \sqrt{\epsilon'}/2. \\
}
The purifications are unitarily equivalent to the state $\ket{\eta},$ up to the specifying unitary $W(p)$ we used in our derivations.
Using this we can bound the overall distance of Eq. (\ref{overall}): 
\EQ{
\Delta \leq \sqrt{\epsilon'} + 4\sqrt{\epsilon''} \leq 2^{-N/4} + 4 \sqrt{ 2^{-N/2-1} + 2^{-3N/4+2} } \leq 5\times 2^{-N/4} 
}
where the last inequality holds for reasonable values of $N$ (but is exponentially decaying either way).

We summarize this as the following theorem:
\TH
The protocol $\pi = (\pi_A, \pi_B),$ with security parameter $N=2K,$ as given in Protocol \ref{prot21}, securely constucts the resource $MRSP_B(4),$ using $N$ calls to the resource $SP(2),$ within $\epsilon,$ with 
\EQ{
 \epsilon \leq 5\times 2^{-N/4}.
}
\HT
We note that the construction of the simulator is also efficient, which matters if the protocol is used in computationally secure settings.

Put all together, we have shown the following sequence of constructions:
\EQ{
SP(2) \rightarrow MRSP_B(4) \rightarrow MSP_B(4) \rightarrow MRSP_B(8) \rightarrow \mathcal{S}_b,
}
where only the first construction is not with zero (but exponentially small) error.

In other words, the capacity to generate the two states $\ket{+},$ $\ket{+_{\pi/2}}$ at random suffices for Alice to securely implement a blind quantum computation protocol. To achieve final error $\epsilon,$ for a blind quantum computing protocol of polynomial size, requiring $M$ gates/qubits, Alice needs to set the security parameter $N = 2\log(M/\epsilon),$ for each call to $SP(2),$ so it is also efficient whenever the computation is polynomial.

In the analysis above, we assumed that the states Alice can generate come from the set $\{\ket{+}, \ket{+_{\pi/2}}\}$,
but the same techniques (and essentially the same protocol) can be used to achieve security for the states of the form  $\{\ket{+}, \ket{+_{\pi/(2K)}}\},$ for any $K$. In other words, the overlap between the two states Alice can generate can be arbitrarily small (or, equivalently arbitrarily large), causing only a logarithmic overhead in the angle (this holds by Eq. (\ref{distance}), and  the small angle approximation for the sinus function).
In such a protocol it will not be guaranteed that the final state ends up in one of the 4 BB84 states (or even the 8 states used in UBQC), but this can be compensated by having Alice send an angle correction, bringing the state to any of the target states. 
Instead of re-writing the entire proof in the more general form, there is a more direct way to see that blindness can be guaranteed even if Alice has access to a device which generates states with arbitrarily small overlap.

\subsubsection{Blindness from states with arbitrarily large overlap}
Here we show that blindness can be achieved by using two states $\ket{+}, \ket{+_\phi},$ where it holds that the angle $\phi$ is smaller than $1/n,$ for any $n$.
The basic idea is to show that for any angle $\phi$, there exists (a smaller angle) $\phi'$ and a unitary $U$ such that
\EQ{
U \ket{+} \ket{0} = \ket{+} \ket{+}\\
U \ket{+_\phi} \ket{0} = \ket{+_{\phi'}} \ket{+_{\phi'}},
}
up to global phase. The sufficient criterion for such a unitary to exist  is that the Gram matrices of the input set $\{\ket{+}\ket{0}, \ket{+_\phi} \ket{0} \}$ and the output set $\{ \ket{+} \ket{+}, \ket{+_{\phi'}} \ket{+_{\phi'}} \}$ coincide \cite{Chefles04}.
This criterion is satisfied when
\EQ{
| \bra{+}  +_\phi \rangle | =  | \bra{+}  +_{\phi'} \rangle |^2 \label{crit1}.
}
In other words, if Alice has access to a device which generates two states $\ket{\phi}, \ket{\psi}$, such that
$ |\bra{\phi} \psi \rangle | = \sqrt{ | \bra{+} +_{\pi/2}\rangle|},$ Alice can run exactly the same protocol as for the two states, where she substitutes each $\ket{+}$ state with $\ket{\phi}\ket{\phi}$ and  $\ket{+_{\pi/2}}$ state with  $\ket{\psi}\ket{\psi}$. She need only instruct Bob to apply the unitary $U$ to each pair, and the security is trivial as Bob could have generated those pairs states from the states  $\ket{+}, \ket{ +_{\pi/2}}$ via an isometry.

This procedure of `halving' the overlap can be iterated, and if iterated $n$ times, we end up with states $\ket{\phi^{n}}, \ket{\psi^{n}}$ such that
\EQ{
 |\bra{\phi^{n}} \psi^{n} \rangle | = | \bra{+} +_{\pi/2}\rangle|^{1/2^{n}},
} the overlap of which approaches unity exponentially quickly.
  
In terms of $CQ$ correlations, this implies that a gadget capable of producing individually asymptotically uncorrelated states 
of the form
\EQ{
\sigma_{AB} = \sum_{b=0}^{1}1/2 \dm{b} \otimes \dm{\psi_b}
}
where $1/2||\dm{\psi_0} - \dm{\psi_1}|| \leq \epsilon,$  
for any $\epsilon,$ still suffices for UBQC.
These states are arbitrarily uncorrelated in the sense that the trace distance between $\sigma_{AB}$ and the closest product state $\sigma_A \otimes \sigma_B$ is at most $\epsilon$.
This, naturally, does not imply that the overall correlations between Alice and Bob can become negligible - Alice will need to utilize a significantly higher number of such low-correlated states to achieve UBQC.
We note that similar techniques to the ones we have applied can be used to show that also highly correlated states, where the overlap between the states Alice can prepare is almost (but not exactly) zero, suffice for constructing $MRSP_B(4)$ and thus run secure blind quantum computation, and indeed, it is likely that protocols can be constructed which achieve security using any two pure states which are neither identical nor orthogonal.
This should be easily provable provided that the angle between the states is not irrational, as in this case, the additive orbit (all states that can be reached) achieved by summing over a large number of such randomly chosen states comprises a finite set (and indeed in the limit we achieve a uniform distribution over that finite set).

In the case of irrational angles, we approach the uniform measure over the real set of the angles, which may require slightly more involved proof techniques. The simplest solution would be to approximate a discrete set, however, in this case errors must be taken into account, and this may be problematic as they accumulate through the protocol. 

The sensitivity to errors also holds true for the protocols we have presented. Any error occurring on Alice's side accumulates additively in the final states. Thus, for every security level desired, one can, in principle, compute the threshold of local errors which can be tolerated, such that the final UBQC computation, if run in a fault-tolerant fashion, still yields correct computation.
Recall that if we require $M$ qubits for the UBQC computation, 
for security level $\epsilon$, we require $O(2\log(M/\epsilon))$ states used in the construction of the resource $MRSP_B(4),$ which is efficient as long as the computation itself is efficient (thus its size is polynomial in the input size). Since the error is additive, the same expression provides a means to compute the local error that can be tolerated: If we can tolerate error $\eta$ per qubit in the UBQC computation, then we require the error $\eta' \in O(\eta(2\log(M/\epsilon))^{-1})$ for each system generated using the $SP(2)$ functionality.

\section{Discussion}
In this work we have studied the types of correlations which are sufficient for achieving blind quantum computation.
In particular, we have identified a class of functionalities which can serve as a substitute for Alice's capacity to generate single qubits in the set $\{ \ket{+_{k\pi/4}} \}_{k=0}^{7}$. Our primary motivation for this was to understand what ultimate purpose the classical-quantum correlations play in blind quantum computing protocols and to analyze whether such correlations can be, asymptotically, be reduced to classical correlations.
The results of this work do not establish the impossibility of a fully classical blind quantum computing scheme - as we have clarified, proving either the existance, or non-existance of such a scheme would have non-trivial consequences in complexity theory, which suggests that coming up with such a protocol (if it exists) will involve more advanced techniques.
Nonetheless, this work does push the boundaries to how restricted (in a certain sense) Alice's devices may be.
Using our constructions, we have provided protocols which achieve security for blind quantum computing (in the full composable sense), where Alice only has the capacity to prepare two states, the overlap of which can be arbitrarily large - leading to arbitrarily small correlations per system.
As a specific example, we show the capacity to prepare the states $\ket{+}, \ket{+_{\pi/2}}$ suffices for the secure construction of the blind quantum computing functionality $\mathcal{S}_b$, with error $\epsilon$ which decays exponentially in the number of states Alice uses. This we further generalize to show that states with arbitrarily large overlap (arbitrarily small difference angle) also suffice for blind quantum computing.

\section*{Acknowledgements}
VD and EK thank Damian Markham, Anthony Leverrier and Joseph Fitzsimons for many discussions which contributed to parts of this paper. VD is particularly indebted to Christopher Portmann for introducing him to the Abstract Cryptography framework and also for invaluable discussions, and comments which helped improve this paper.
Parts of this work were done while VD was supported by the EPSRC (grant EP/G009821/1), EPSRC Doctoral Fellowship, and by the Austrian Science Fund (FWF) through the SFB FoQuS F 4012. EK acknowledges funding through EPSRC grants EP/N003829/1 and EP/M013243/1.
 \bibliographystyle{unsrt}
\bibliography{classical,quantum,thesis}
\end{document}